\documentclass{emulateapj} 
\usepackage{graphics,graphicx,lscape}
\usepackage{apjfonts}

\def \arcmin {\hbox{$^\prime$}} 
\def \arcsec {\hbox{$^{\prime\prime}$}} 
 
\def\spose#1{\hbox to 0pt{#1\hss}} 
\def\ltsim{$\mathrel{\spose{\lower 3pt\hbox{$\sim$}} 
        \raise 2.0pt\hbox{$<$}}$\thinspace} 
\def\gtsim{$\mathrel{\spose{\lower 3pt\hbox{$\sim$}} 
        \raise 2.0pt\hbox{$>$}}$\thinspace} 
\def \msun {${\rm M_\odot}$}

\newcommand{\rfive}{${\rm r_{500}}$}

\newcommand{\chandra }{{\em Chandra}}

\newcommand{\ciao }{{\em CIAO}} 
\newcommand{\caldb }{{\em Caldb}} 
\newcommand{\heasoft }{{\em Heasoft}}

\newcommand{\rosat }{{\em ROSAT}}

\newcommand{\heasarc }{{\em HEASARC}\footnote{ftp://legacy.gsfc.nasa.gov}}

\newcommand{\thin}{\thinspace}
\newcommand{\ptwo}{${\rm P_2/P_0}$}
\newcommand{\pthree}{${\rm P_3/P_0}$}

\shorttitle{Rotation and Turbulence in Galaxy Clusters}
\shortauthors{Fang et al}

\makeatletter 

\received{} 
\revised{}
\accepted{}

\shorttitle{} 
\shortauthors{Fang et al.}

\begin{document}

\title{Rotation and Turbulence of the Hot ICM in Galaxy Clusters}

\author{Taotao~Fang\altaffilmark{1}, Philip~Humphrey\altaffilmark{1}, and 
  David~Buote\altaffilmark{1}} 
  
\altaffiltext{1}{Department of Physics \& Astronomy, 4152 Frederick Reines Hall
, University of California, Irvine, CA 92697; fangt@uci.edu}
   
\begin{abstract}
Cosmological simulations of galaxy clusters typically find that the
weight of a cluster at a given radius is not balanced entirely by the
thermal gas pressure of the hot ICM, with theoretical studies
emphasizing the role of random turbulent motions to provide the
necessary additional pressure support. Using a set of high-resolution,
hydrodynamical simulations of galaxy clusters that include radiative
cooling and star formation and are formed in a Cold Dark
Matter universe, we find instead that in the most relaxed clusters
rotational support exceeds that from random turbulent motions for
radii, $0.1- 0.5\, r_{500}$, while at larger radii, out to
$0.8r_{500}$, they remain comparable.  We also find the X-ray images of
the ICM flatten prominently over a wide radial range, $0.1-0.4\, r_{500}$.  When
compared to the average ellipticity profile of the observed X-ray
images computed for 9 relaxed nearby clusters, we find that the
observed clusters are much rounder than the relaxed CDM clusters
within $\approx 0.4r_{500}$.  Moreover, while the observed clusters
display an average ellipticity profile that does not vary
significantly with radius, the ellipticity of the relaxed CDM clusters
declines markedly with increasing radius, suggesting that the ICM of
the observed clusters rotates less rapidly than that of the relaxed
CDM clusters out to $\approx 0.6r_{500}$. When these results are
compared to those obtained from a simulation without radiative
cooling, we find a cluster ellipticity profile in much better
agreement with the observations, implying that over-cooling has a
substantial impact on the gas dynamics and morphology out to larger
radii than previously recognized. It also suggests that the
10\%-20\% systematic errors from non-thermal gas pressure support
reported for simulated cluster masses, obtained from fitting simulated
X-ray data over large radial ranges within $r_{500}$, may need to be
revised downward. These results demonstrate the utility of X-ray
ellipticity profiles as a probe of ICM rotation and over-cooling which should be used
to constrain future cosmological cluster simulations.
\end{abstract}


\section{Introduction} \label{intro}

Clusters of galaxies, the largest virialized structures in the
universe, play a critical role in understanding the formation and
evolution of large scale structure, and determining fundamental
cosmological parameters. It is now well-known that the dark matter,
the dominant component, makes up about 80\% of the total gravitational
mass of a typical galaxy cluster. The baryonic matter in galaxies
accounts for about 3 --- 5\% of the total mass. The remaining mass is
made up by hot gas distributed between galaxies, with temperatures
over $10^7$ K. This hot gas, called the ``intracluster medium'' (ICM),
emits X-rays through thermal bremsstrahlung radiation, and makes
galaxy clusters the second most luminous X-ray sources in the
universe, after quasars. In the era of precision cosmology, it becomes
imperative to measure accurately cluster properties, such as its
gravitational mass, to a precision of a few percent.

The X-ray measurement of the density and temperature profiles of the
hot ICM provides one of the most accurate methods to determine the
cluster total gravitational mass. For clusters that have not been
disturbed recently by a major merger, the hot ICM should obey
hydrostatic equilibrium to a good approximation, representing a
balance between the gravitational force and thermal gas
pressure.  For the special case of spherical symmetry the
equation of hydrostatic equilibrium may be written,
\begin{equation}
\frac{dP_g}{dr} = -\rho_g\frac{GM(r)}{r^2}, \label{eqn.he}
\end{equation}
where $\rho_g$ is the gas density, $P_g$ is the thermal gas pressure,
and $M(r)$ is the mass enclosed within radius $r$. Applications of the
X-ray method to constrain cosmological parameters include, e.g.,
measurements of the gas mass fraction (Allen et al.~2007, and
references therein), virial mass function (e.g., Stanek et al.~2006,
and references therein) and the relationship between the virial
concentration and mass (Buote et al.~2007).

It is necessary to use cosmological hydrodynamical N-body simulations
to test the key assumptions underlying equation [\ref{eqn.he}]; i.e.,
spherical symmetry and negligible non-thermal pressure support of the
hot ICM (e.g., Evrard et al.~1996; Thomas et al.~1998).  Early studies showed that it is possible to recover the
cluster mass within a factor of $\sim 2$ with X-ray telescopes such as
{\sl EINSTEIN} and {\sl EXOSAT} (e.g., Tsai et al.~1994), where
isothermal gas had to be assumed in the mock X-ray analysis. Most
recent studies, which involve mock observations with the most advanced
telescopes such as {\sl Chandra}, allow the radial gas temperature
profile to be measured, and use simulations that include a wider range
of physical processes. They find that the global cluster mass,
typically computed within a radius of fixed over-density (e.g.,
$r_{500}$), can be recovered more accurately, with systematic errors
typically ranging from 5\% to 20\% (see, e.g., Nagai et al.~2007).

\begin{figure*}[t]
\begin{center}
\includegraphics[angle=0]{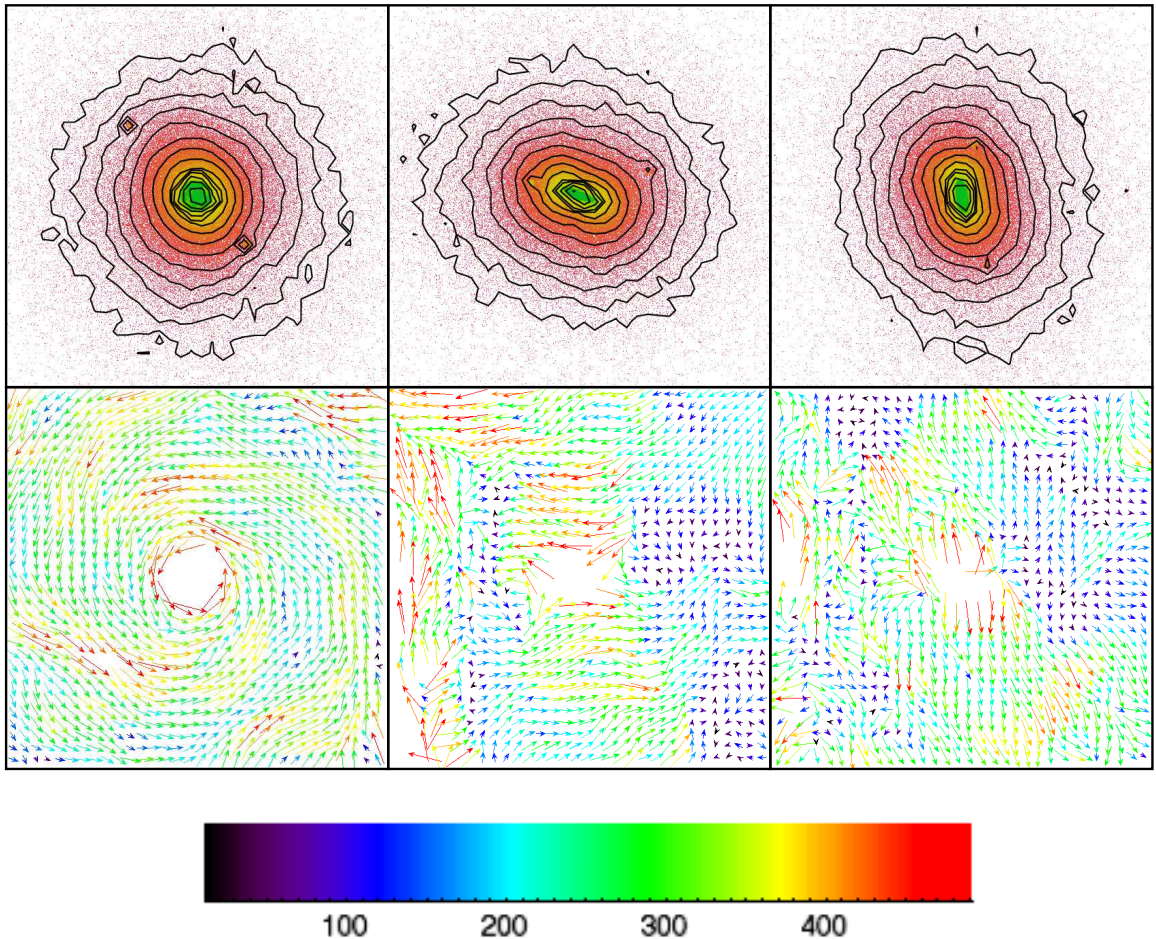}
\end{center}
\caption{Top panel: mock {\sl Chandra} X-ray flux maps of CL7 from $x$
projection (left), $y$ projection (middle), and $z$ projection
(right). Bottom panel: slice of the velocity field in the center of
cluster CL7, $y-z$ plane (left), $x-z$ plane (middle), $x-y$ plane
(right). The bottom color bar is for the velocity field, and in units
of $\rm km\ s^{-1}$. Each box has a size of $1h^{-1}\times1h^{-1}\rm\ Mpc^2$.}
\label{image}
\end{figure*}

\begin{figure*}[t]
\begin{center}
\includegraphics[angle=0]{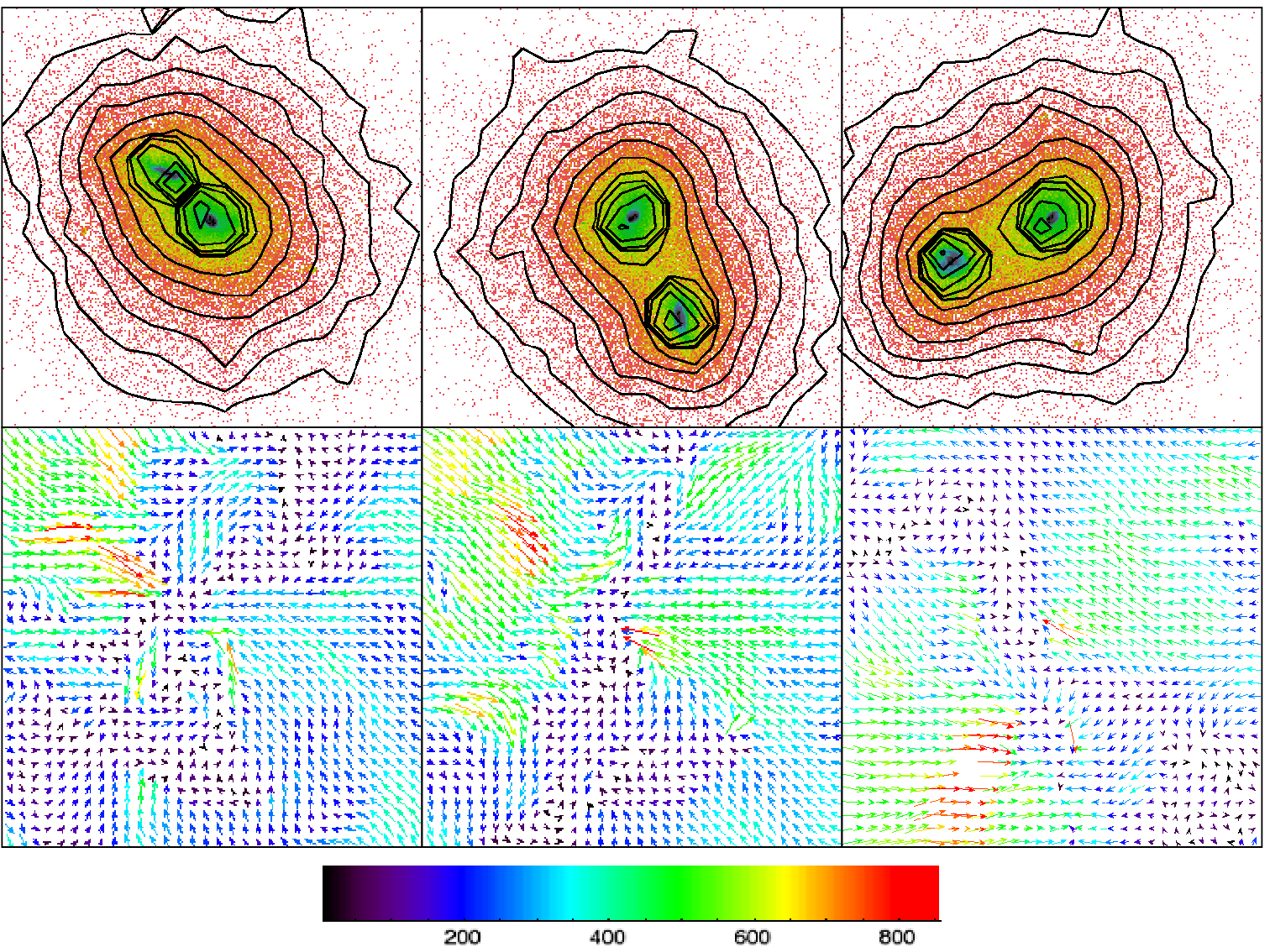}
\end{center}
\caption{Top panel: mock {\sl Chandra} X-ray flux maps of CL11 from $x$
projection (left), $y$ projection (middle), and $z$ projection
(right). Bottom panel: slice of the velocity field in the center of
cluster CL11, $y-z$ plane (left), $x-z$ plane (middle), $x-y$ plane
(right). The bottom color bar is for the velocity field, and in units
of $\rm km\ s^{-1}$. Each box has a size of $1h^{-1}\times1h^{-1}\rm\ Mpc^2$.}
\label{image11}
\end{figure*}

Over the past decade simulators have emphasized how much additional
pressure support from random turbulent gas motion contributes to
systematic errors in the X-ray method (e.g., Bryan \& Norman~1998; Kay
et al.~2004; Faltenbacher et al.\ 2005; Dolag et al.~2005; Rasia et
al.~2006; Hallman et al.~2006; Vazza et al.~2006; Nagai et
al.~2007; Brunetti \& Lazarian~2007).  It has also been noticed in such high-resolution,
hydrodynamic simulations that while the velocity dispersion tensor of
the ICM is approximately isotropic in the outskirts of clusters, it
becomes increasingly tangential at smaller radii, especially for the
most relaxed systems (Rasia, Tormen, \& Moscardini 2004; Lau et al.~2008, in preparation); Recently Lau et al.\ (2008)
have shown the ICM velocity dispersion becomes increasingly
tangentially anisotropic as one moves inward from $r_{500}$, such that
for their relaxed clusters the anisotropy parameter falls to
$\beta\approx -0.3$ near $0.2r_{500}$.  These authors consider the
non-thermal pressure support of this random turbulent motion (both
radial and tangential) on estimates of $M_{500}$ assuming hydrostatic
equilibrium. However, rotational support of the gas, and its
observable signatures, is not addressed.

In this paper, we show that for the relaxed clusters studied by Lau et
al. (2008; Nagai et al.\ 2007) support of the ICM from rotational and
streaming motions is comparable to the support from the random
turbulent pressure out to $\approx 0.8 r_{500}$.  While the overall
magnitude of the rotational motion ($\sim$ a few hundred $\rm km\
s^{-1}$) is not large enough to be detected directly through Doppler
shifts of emission lines in X-ray spectra, even with the most advanced
X-ray telescopes we have today (e.g., Sunyaev et al.~2003; Inogamov \&
Sunyaev~2003; Schuecker
et al.~2004;Br{\"u}ggen et al.~2005; Pawl et al.~2005; Chepurnov \& Lazarian~2006; but
see Dupke \& Bregman~2006 for a recent mmeasurement), simulated clusters with large-scale rotation are
significantly flat, and this translates to observable large ellipticities of the X-ray
isophotes. By comparing the ellipticities of the X-ray isophotes of the
relaxed simulated clusters to those of nine observed clusters, we show
that the observed clusters are, on average, much rounder and have a
distinctly different radial variation in ellipticity.  This
demonstrates the utility of X-ray ellipticity profiles as a constraint
for future cosmological cluster simulations.

This paper is organized as follows. In \S2 we study the importance of
ICM rotation in the 16 simulated clusters of Lau et al. (2008; Nagai
et al.\ 2007), focusing our discussion using the examples of one
relaxed and one disturbed cluster.  In \S3 we present ellipticity
profiles of nine clusters obtained from X-ray observations with {\sl
Chandra} and {\sl ROSAT} and compare them to the simulated
clusters. The last section is devoted to summary and discussion.

\section{Non-thermal Gas Motion in the Simulated Clusters}

\subsection{Simulation Data}

We use a set of 16 high-resolution, cosmological hydrodynamic
simulations of cluster-sized systems in a flat $\Lambda CDM$ model:
$\Omega_m = 0.3, \Omega_{\Lambda}=0.7, \Omega_b = 0.043$ and $\sigma_8
= 0.9$ (the power spectrum normalization at 8$h^{-1}$ Mpc
scale). These cluster-sized simulations include collisionless dynamics
of dark matter, star and intracluster gas. They also include several
critical physical processes such as radiative cooling, star formation and metal
enrichment (CSF simulation). For comparison we also analyze one adiabatic cluster
simulation, i.e., no radiative cooling and star formation (NC simulation). We refer readers to Nagai et al.~(2007) for details, and
provide a brief summary here. We adopt a Hubble constant of $H_0 =
100h\rm\ km\ s^{-1}Mpc^{-1}$ where $h=0.7$ throughout the paper.

The simulations use the Adaptive Refinement Tree (ART) N-body +
gasdynamics code (Klypin et al.~2001; Kravtsov et al.~2002, 2006; Nagai
et al.~2007), and run in two computational boxes: $120h^{-1}$ Mpc for
CL101--107, and $80h^{-1}$ Mpc for the rest. The ART is a
shock-capturing code that is particularly suited for capturing shocks
and turbulent motion, and the use of adaptive mesh refinement also
allows us to resolve the detailed cluster structure and the internal
gas flows in clusters. These simulations achieve a dynamic range of
$\sim 5\times 10^{13} $--- $10^{15}h^{-1}\ M_{\odot}$, and a peak
resolution of $\sim 6h^{-1}$ kpc, which is well-matched to the spatial
resolution of Chandra X-ray observations of nearby galaxy
clusters. Table~1 of Nagai et al.~2007 shows the typical properties of
simulated clusters, all the parameters are computed from simulations
within a region with an over-density of 500, with respect to the
critical density of the universe. These clusters have masses ranging
between $\sim 3\times 10^{13}h^{-1}\rm\ M_{\odot}$ and
$10^{15}h^{-1}\rm\ M_{\odot}$, and temperatures between 1 keV and 8.7
keV.

\begin{figure*}[t]
\begin{center}
\includegraphics[angle=90,scale=0.37]{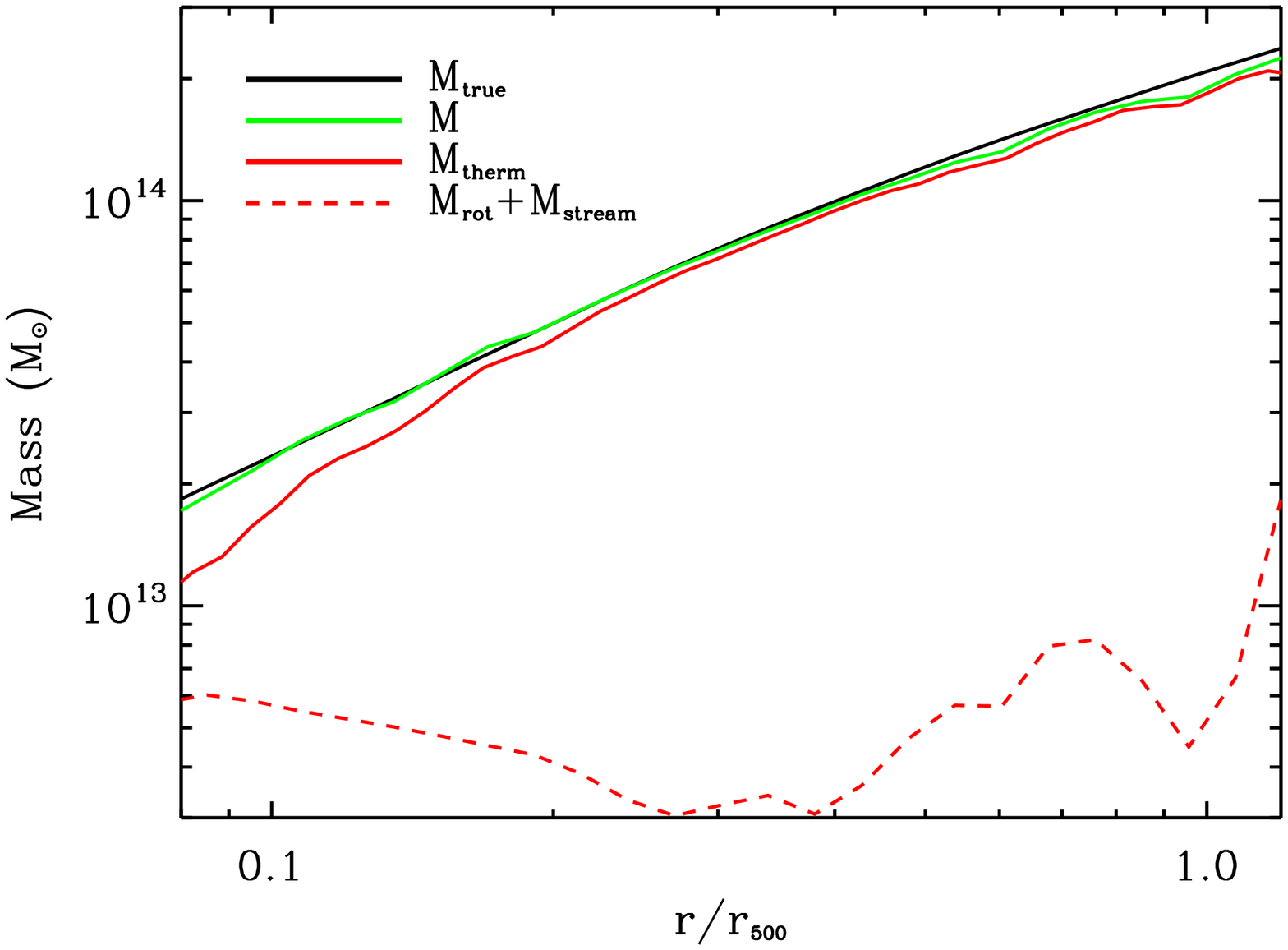}
\includegraphics[angle=90,scale=0.37]{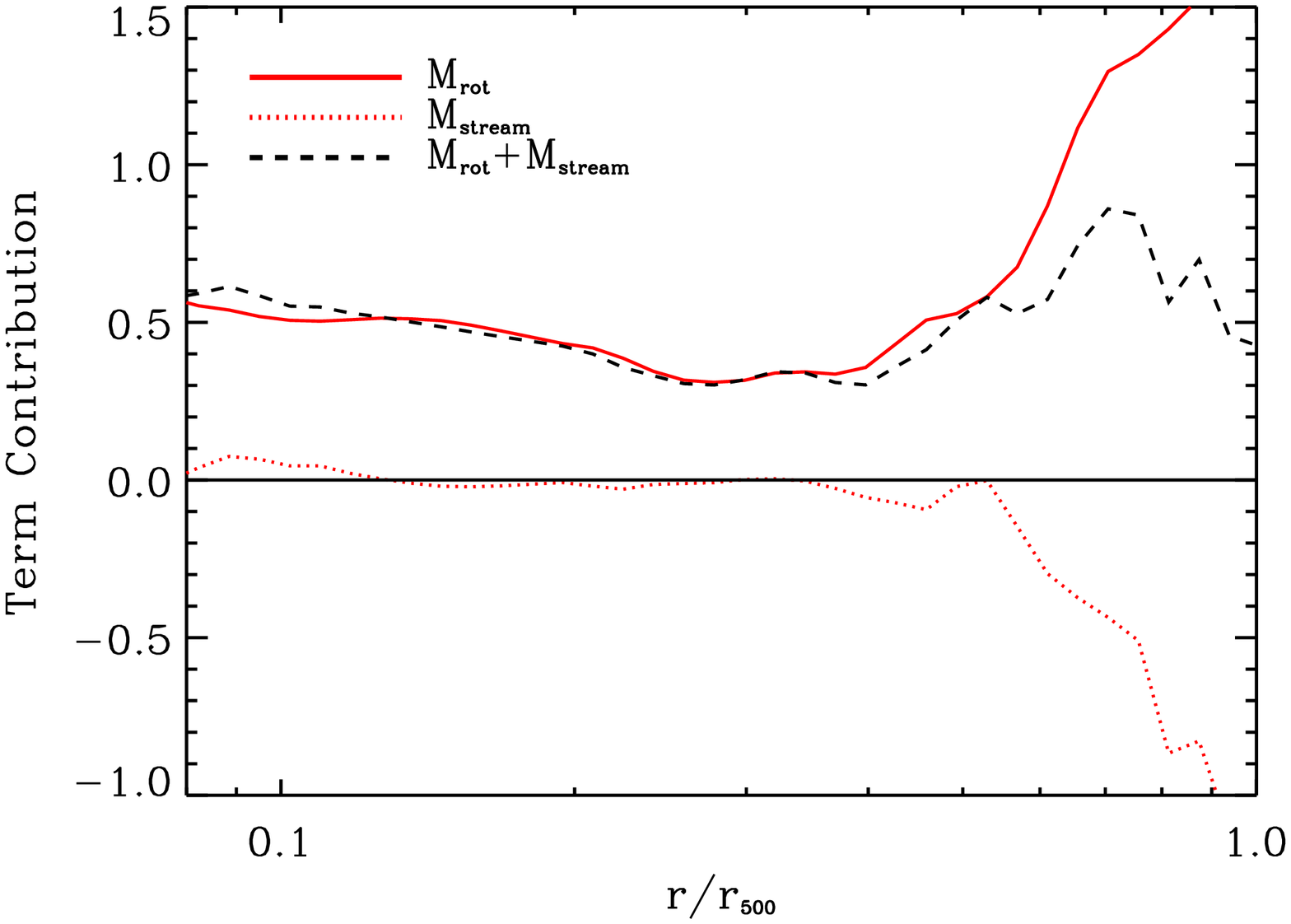}
\end{center}
\caption{({\sl Left Panel}) Mass profiles of the relaxed simulated cluster CL7.
The ``true'' mass $M$ obtained by directly counting the masses of the
gas and dark matter particles from the simulations is given by the
dark solid line. The other lines represent the different mass
components obtained by substituting Euler's equation into Gauss's Law
(see \S \ref{method}): the component from thermal gas pressure,
$M_{therm}$ (red solid line), rotation and streaming motion,
$M_{rot}+M_{stream}$ (red dashed line), and their sum,
$M_{therm}+M_{rot}+M_{stream}$ (green solid line). ({\sl Right Panel})
Separate contributions of $M_{rot}$ (red solid line) and $M_{stream}$
(red dotted line) to the mass profile, in units of $10^{13}\rm\
M_{\odot}$. The dashed black line is the total contribution,
$M_{rot}+M_{stream}$.}
\label{m7}
\end{figure*}

The simulated clusters are projected along three orthogonal directions
${x,y,z}$, and X-ray flux maps are created for each projection.
(These coordinate axes reflect the orientation of the entire
simulation box and are not adjusted to match the principle axes of
each cluster.) We make the mock {\sl Chandra} images by convolving the
emission spectrum with the response of the {\sl Chandra}
front-illuminated CCDs and with an exposure of 100 $ksec$, a typical
exposure for {\sl Chandra} observations. We show the images and
velocity structures of a relaxed cluster (CL7) in Figure~\ref{image}
and a disturbed cluster (CL11) in Figure~\ref{image11}. CL7 is judged
to be the most relaxed cluster in the simulation sample, with a
gravitational mass of $M_{500}=2.01\times10^{14}\rm\ M_{\odot}$ and a
size of $r_{500}=0.891$ kpc. Here $r_{500}$ is defined as a radius of
500 times the mean density of the universe, $M_{500}$ is the total
gravitational mass within $r_{500}$. In contrast, CL11 is a very
disturbed system with $M_{500}=1.29\times10^{14}\rm\ M_{\odot}$ and
$r_{500}=0.767$ kpc.

The top panel of Figure~\ref{image} shows the projected X-ray emission
maps of CL7, in the three orthogonal directions ${x,y,z}$ from left to
right, respectively. The X-ray contours (solid lines) are separated by
a factor of $\sim 2$ in surface brightness, with the highest contour
at a level of $\sim 500$ photons per pixel. The box size of the image
is $1\ h^{-1}$ Mpc. The cluster has a very regular appearance in all
projections. However, the X-ray isophotes display significant
elongation in the $y$ and $z$ projections compared to the much rounder
isophotes of the $x$ projection. 

The bottom panel of the Figure~\ref{image} shows the two dimensional
velocity field in the central slices of the simulated cluster CL7. The
left plot is along the $y-z$ plane, the middle plot is along the $x-z$
plane, and the right is the $x-y$ plane. The velocity vector is
color-coded between 0 and 500 $\rm km\ s^{-1}$, with the magnitude
showing in the bottom color bar. The most striking feature is that
while in both $x-z$ and $x-y$ planes the gas is moving largely
randomly, in $y-z$ plane the gas shows a very regular,
counterclockwise, rotational motion with velocities between $200 -
400\rm\ km\ s^{-1}$.

In contrast, CL11 is a merging cluster with violent gas motion
(Figure~\ref{image11}). Merging in the central region creates large
scale gas motions on the order of 1,000 $\rm\ km\ s^{-1}$. Such merging
activity is clearly visible in the velocity fields shown in the $y-z$
plane and $x-z$ plane: gas is moving from the north-west and
south-east directions toward the center. The X-ray images in the three
projections are also distorted. Unlike CL7 ordered rotational motion
is unimportant in all projections.

In the following subsections we compare the contribution of rotation
(and other streaming motion) to that of random turbulent motion in
supporting the weight of the ICM of the simulated clusters.

\subsection{Analysis Method} \label{method}

The total cluster mass $(M)$ enclosed within a surface $\mathbf{S}$ is
given by Gauss's Law,
\begin{equation}
\nonumber
M  =  \frac{1}{4\pi G}\int \mathbf{\nabla} \Phi \cdot d^2 \mathbf{S}, 
\end{equation}
where $\Phi$ is the gravitational potential and $G$ is the gravitational constant.
Assuming the ICM is a steady state, inviscid, collisional fluid we may use
Euler's equation to eliminate $\Phi$ in favor of the gas pressure and
terms involving the velocity components of the ICM,
\begin{eqnarray}
\nonumber
M  =  \frac{1}{4\pi G}\int
\left[-\frac{1}{\rho_g}\mathbf{\nabla}
P_g-\left(\mathbf{v}\cdot\nabla\right)\mathbf{v}\right] \cdot d^2 \mathbf{S},
\end{eqnarray}
where $\rho_g$ is the gas density and $P_g$ represents the pressure
arising from gas motions having a random (i.e., Maxwellian) velocity
distribution.

We rewrite the previous equation as a sum of four terms, 
\begin{equation}
M = M_{therm} + M_{turb} + M_{rot} + M_{stream}.
\end{equation}
The first two terms on the r.h.s.,
\begin{eqnarray}
M_{therm} & = & -\frac{1}{4\pi G}\int
\left(\frac{1}{\rho_g}\mathbf{\nabla} P_{therm}\right)\cdot d^2\mathbf{S} \\
M_{turb} & = & -\frac{1}{4\pi G}\int
\left(\frac{1}{\rho_g}\mathbf{\nabla} P_{turb}\right)\cdot d^2\mathbf{S},
\end{eqnarray}
represent pressure support from random gas motions. Here $P_{therm} =
kT/\mu m_p$ is the thermal gas pressure, where $k$ is the Boltzmann
constant, $T$ is temperature, $m_p$ is the proton mass, and $\mu$ is
the atomic weight. The quantity $P_{turb}$ is the pressure from random
turbulent motions. The last two terms on the r.h.s.,
\begin{eqnarray}
M_{rot} & = & \frac{1}{4\pi G}\int
\left(\frac{\upsilon^2_{\theta}+\upsilon^2_{\phi}}{r}\right)\cdot d^2\mathbf{S}\\
M_{stream} & = & -\frac{1}{4\pi G}\int
\left(\upsilon_r\frac{\partial \upsilon_r}{\partial r} + \frac{\upsilon_{\theta}}{r}\frac{\partial
  \upsilon_r}{\partial \theta} + \frac{\upsilon_{\phi}}{r\sin \theta}\frac{\partial
  \upsilon_r}{\partial \phi}\right)\cdot d^2\mathbf{S},
\end{eqnarray}
follow by evaluating $\left(\mathbf{v}\cdot\nabla\right)\mathbf{v}$ in
spherical coordinates (Binney \& Tremaine~1987). Here $M_{rot}$ is the contribution from rotational
motion while $M_{stream}$ includes the contributions from other
streaming motion.

\begin{figure*}[t]
\begin{center}
\includegraphics[angle=90,scale=0.37]{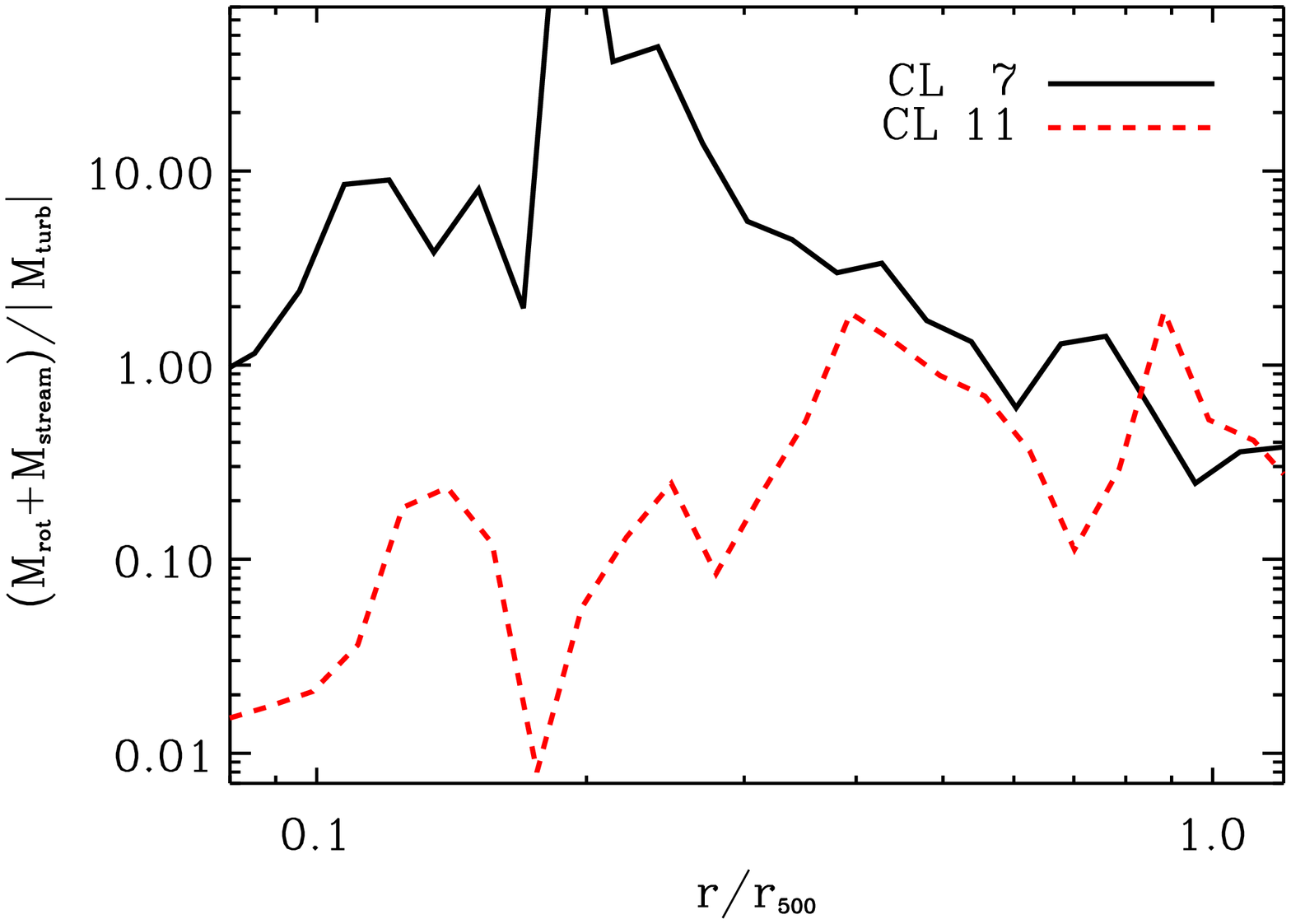}
\includegraphics[angle=90,scale=0.37]{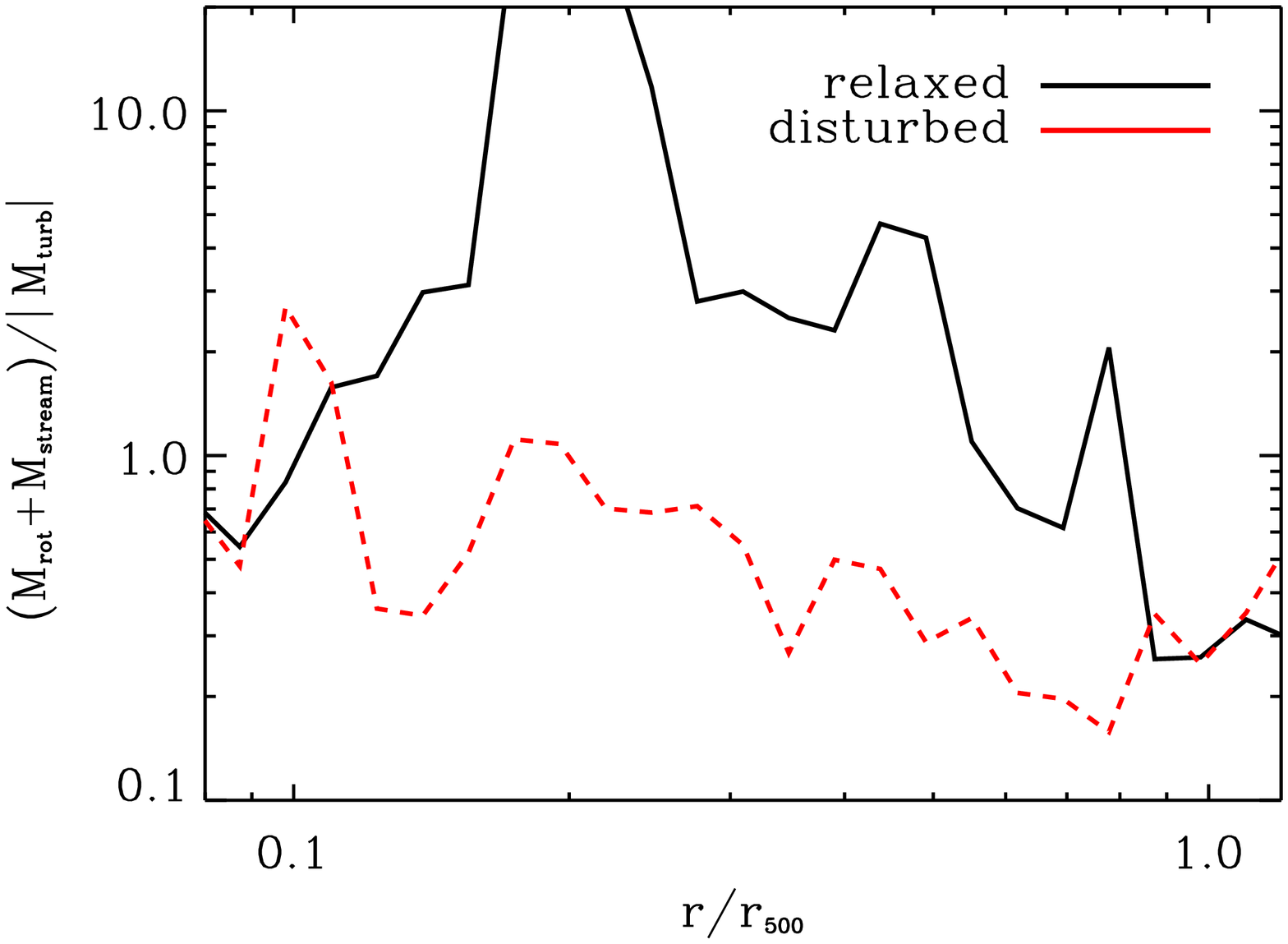}
\end{center}
\caption{({\sl Left Panel}) Relative contribution of rotation and other streaming motion to
the (absolute value of the) turbulent motion in CL7 (solid black line)
and CL11 (dashed red line). ({\sl Right Panel}) The same quantity is
plotted by averaging the results for all of the seven relaxed clusters (solid
black line) and all of the nine disturbed clusters (dashed red line) in the
simulation.}
\label{v7}
\end{figure*}

The combination $M_{rot} + M_{stream}$ represents the total contribution to
the gravitational support from non-random gas motions. In this paper
we are especially concerned with assessing the relative importance of
non-random gas motions to that of the random turbulent
motion. However, we also find it useful to separate the non-random
contribution into the terms $M_{rot}$ and $M_{stream}$. For a cluster
supported mostly by rotation $M_{rot}$ will dominate $M_{stream}$.  For a
non-rotating cluster with substantial non-random motions, $M_{rot}$ and
$M_{stream}$ will be of similar magnitude (but will not necessarily
cancel). Hence, in our discussion below we always consider the total
$M_{rot} + M_{stream}$ when assessing the relevance of non-random gas
motions, but we use $M_{rot}$ ($\gg M_{stream}$) as a rotation proxy
particularly for the relaxed clusters.

In what follows we choose the surface $\mathbf{S}$ to be that of a
sphere, but we emphasize that we do not assume the cluster is
spherical and use the full three-dimensional ICM properties produced
by the simulation to evaluate the integrals.  Furthermore, we do not
explicitly calculate $M_{turb}$ since it depends critically on the
sizes, velocities, temperatures, and densities of the coherent clumps
that move randomly within the ICM. Instead, we infer its magnitude
from $M_{turb} = M - M_{therm} - M_{rot} - M_{stream}$, where $M$ is
the total ``true'' mass of the gas and dark matter computed directly from the
particles in the simulation.

To compute the mass profile associated with both the ``true'' mass and
the mass components based on Eq.[4--7], we make use of the grid data
that is generated from the raw data of the adaptive mesh refinement
simulation. Specifically, the raw data is interpolated onto to a
uniform $256^3$ cube grid in a $2h^{-1}\times2h^{-1}\times2h^{-1}$
Mpc$^3$ box, centered on the simulated cluster. The spatial resolution
therefore is 7.8$h^{-1}$ kpc.

We divide the simulation box into a number of radial bins, and compute
each mass component. At each radial bin, the entire surface is divided
into small cells along $\theta$ and $\phi$ directions, over which the
integration is performed. Specifically, in each cell we throw test
particles randomly, we then compute the mean density, thermal
pressure, and velocity by averaging all the test particles. Again,
since we use Gauss's Law and compute the integration in three
dimensions, our analysis is entirely general with regard to the cluster
geometry.

\subsection{Mass Profiles}

\begin{figure*}[t]
\begin{center}
\includegraphics[angle=90,scale=0.37]{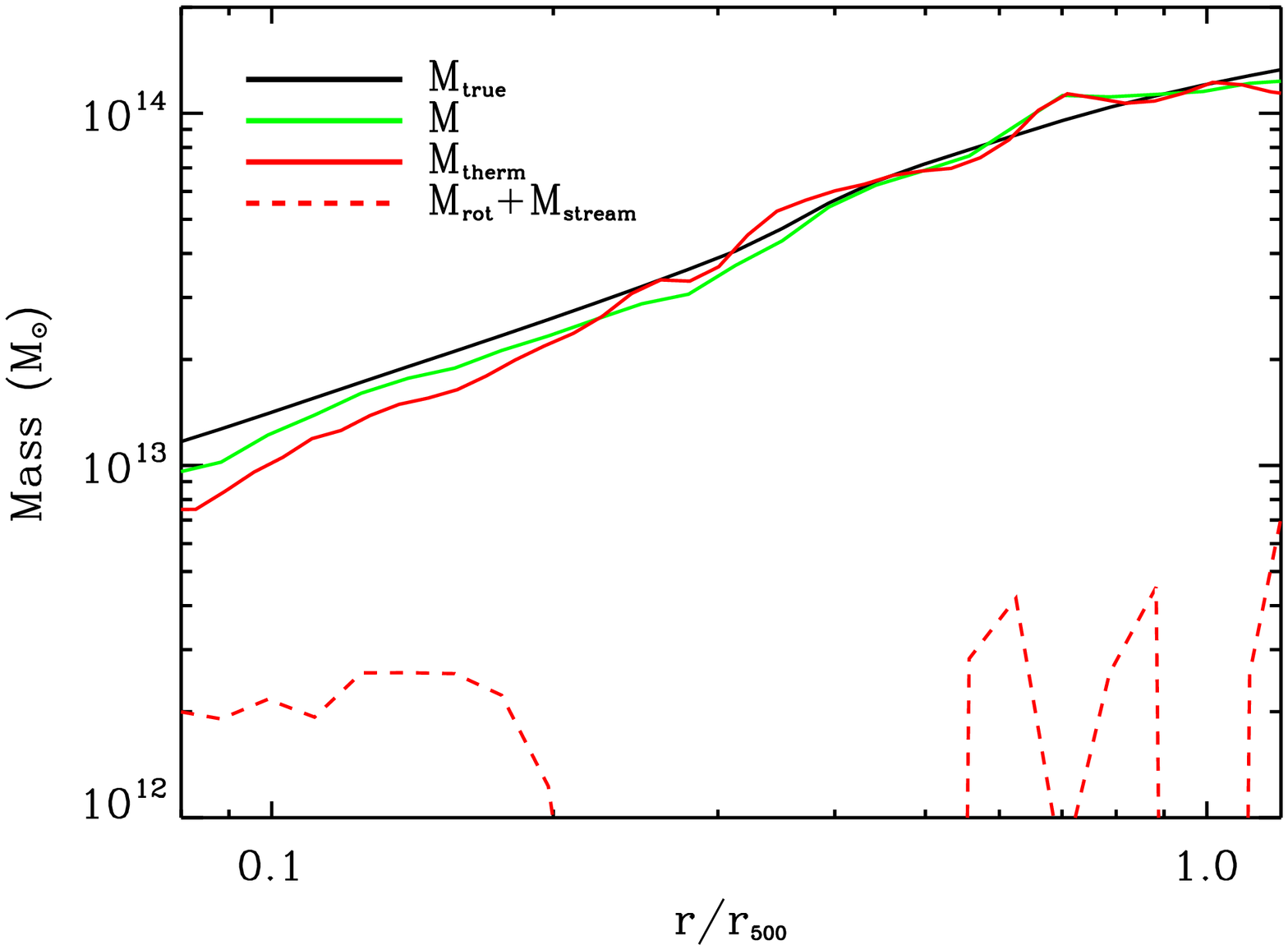}
\includegraphics[angle=90,scale=0.37]{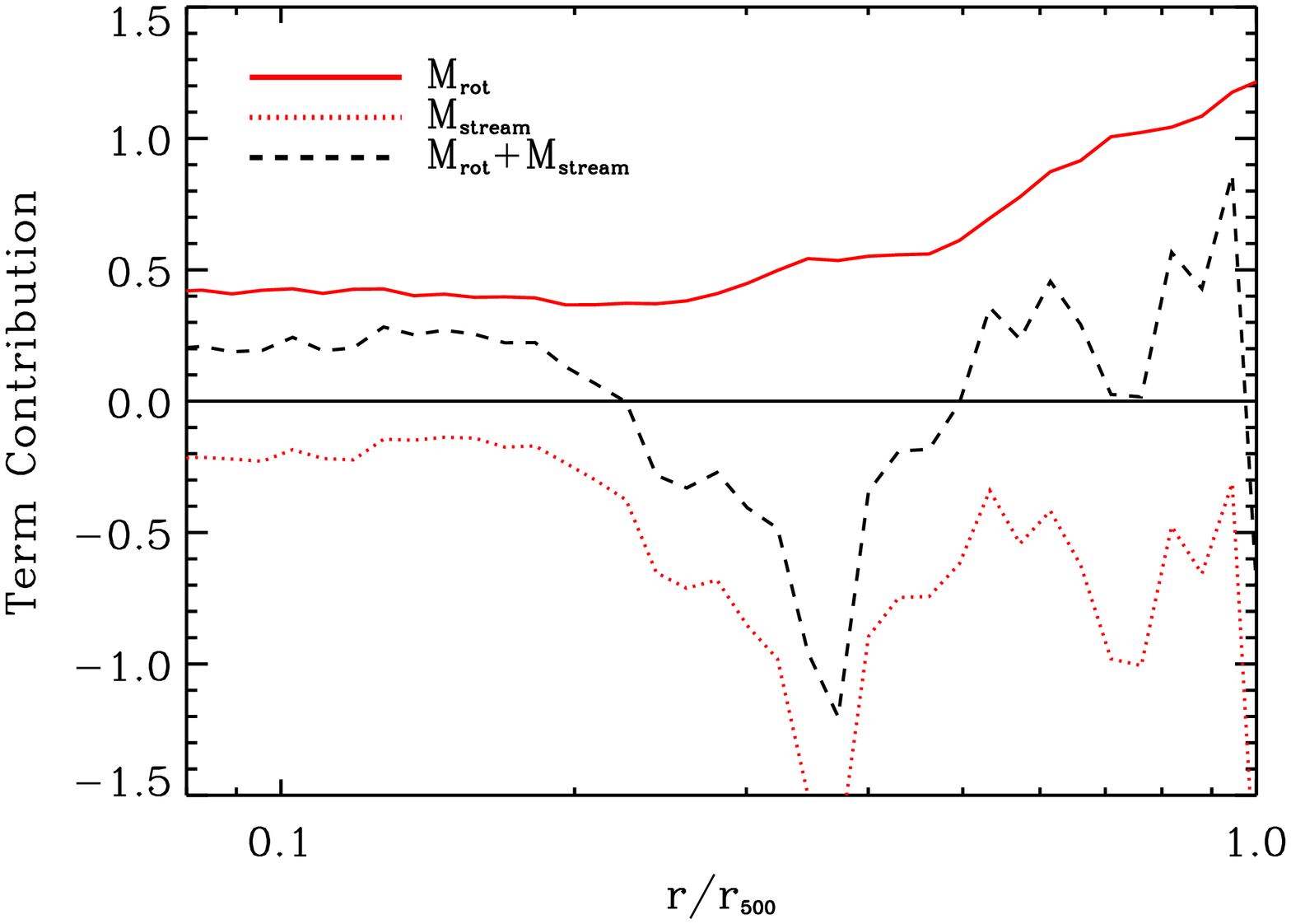}
\end{center}
\caption{Same as Figure \ref{m7} but for the disturbed simulated cluster CL11.}
\label{m11}
\end{figure*}

In the left panel of Figure~\ref{m7} we show the ``true'' mass profile
of the relaxed cluster CL7 obtained by directly counting the dark
matter and mass particles from the simulation.  Also displayed are
various combinations of $M_{therm}$, $M_{rot}$, and $M_{stream}$, that
result from substituting Euler's equation into Gauss's Law as
discussed in the previous section. We consider seriously the gas
properties of the simulated clusters down to a radius of $\approx
0.08r_{500}$. Below this radius the well-known effect of over-cooling
produces unphysical gas density and temperature profiles.

Over most of the cluster region displayed, $M_{therm}$ systematically
underestimates the ``true'' mass by 5\%-10\%. Rotational and other
streaming motions provide most of the non-thermal support of the
cluster weight over a large range of radii. The combination of
$M_{therm}+M_{rot}+M_{stream}$ (green solid line in the
Figure~\ref{m7}) is within a few percent of the ``true'' mass out to
$\approx 0.5r_{500}$. For these radii, $M_{rot} \gg M_{stream}$ (see
right panel of Figure~\ref{m7}), indicating that rotation is the most
important streaming motion, consistent with the visual impression of
the velocity field shown in Figure \ref{image}.

We provide a direct comparison of rotation and streaming motions to
random turbulent motions in Figure \ref{v7}. (Note we take the
absolute value of $M_{turb}$ since it fluctuates about zero when its
magnitude is small.)  Over approximately $0.1-0.5\, r_{500}$,
$M_{rot}+M_{stream}$ generally dominates $M_{turb}$ and remains
comparable to $M_{turb}$ all the way out to $\approx
0.8r_{500}$. Since, as noted above, $M_{rot} \gg M_{stream}$ within
$\approx 0.5r_{500}$, it follows that $M_{rot}$ dominates the
non-thermal support of the cluster weight in that region.

In contrast, for the disturbed cluster CL11 $M_{turb} \ga
M_{rot}+M_{stream}$ over all radii shown (Figures \ref{m7} and
\ref{v7}). Rotational motion does not dominate,
consistent with the lack of circular motion noted from visual
inspection of the velocity field shown in Figure \ref{image11}.
Despite the differences between CL7 and CL11, it is interesting to
note that for CL11 $M_{therm}$ typically lies within $\pm 10\%$ of the
``true'' mass. This is of similar magnitude to CL7, except that CL7
only underestimates the ``true'' mass.

The basic results for CL7 and CL11 apply generally to the other
relaxed and disturbed clusters in the simulation. In particular, we
show in the right panel of Figure \ref{v7} the relative contribution
of rotation and other streaming motion to the turbulent motion by
averaging the results for all of the relaxed clusters (solid black
line) and all of the disturbed clusters (dashed red line) in the
simulation. (We have used the same clusters designated as ``relaxed''
or ``disturbed'' as Nagai et al.~2007). While there is substantial
scatter (not shown) about these average profiles, particularly for the
disturbed systems, $M_{rot}+M_{stream}$ generally dominates $M_{turb}$
over approximately $0.1-0.5\, r_{500}$ and remains comparable to
$M_{turb}$ all the way out to $\approx 0.8r_{500}$. In the disturbed
clusters, $M_{turb}$ exceeds the rotational and streaming terms over
most of the region within $r_{500}$.

\subsection{Adiabatic Simulation}

We also analyze a simulation that starts from the same initial conditions
as those adopted in CL7. However, in this simulation, the ICM of the
cluster -- called ``CL7a'' -- evolves adiabatically, i.e.,
without radiative cooling. The major difference between these two sets of simulations is that the CSF clusters suffer the so-called "over-cooling" problem (e.g., Balogh et al.~2001): cooling can produce  a substantial amount of cold gas that is likely a few times higher than what are observed in real clusters (Gnedin et al.~2004).

\begin{figure*}[t]
\begin{center}
\includegraphics[angle=90,scale=0.35]{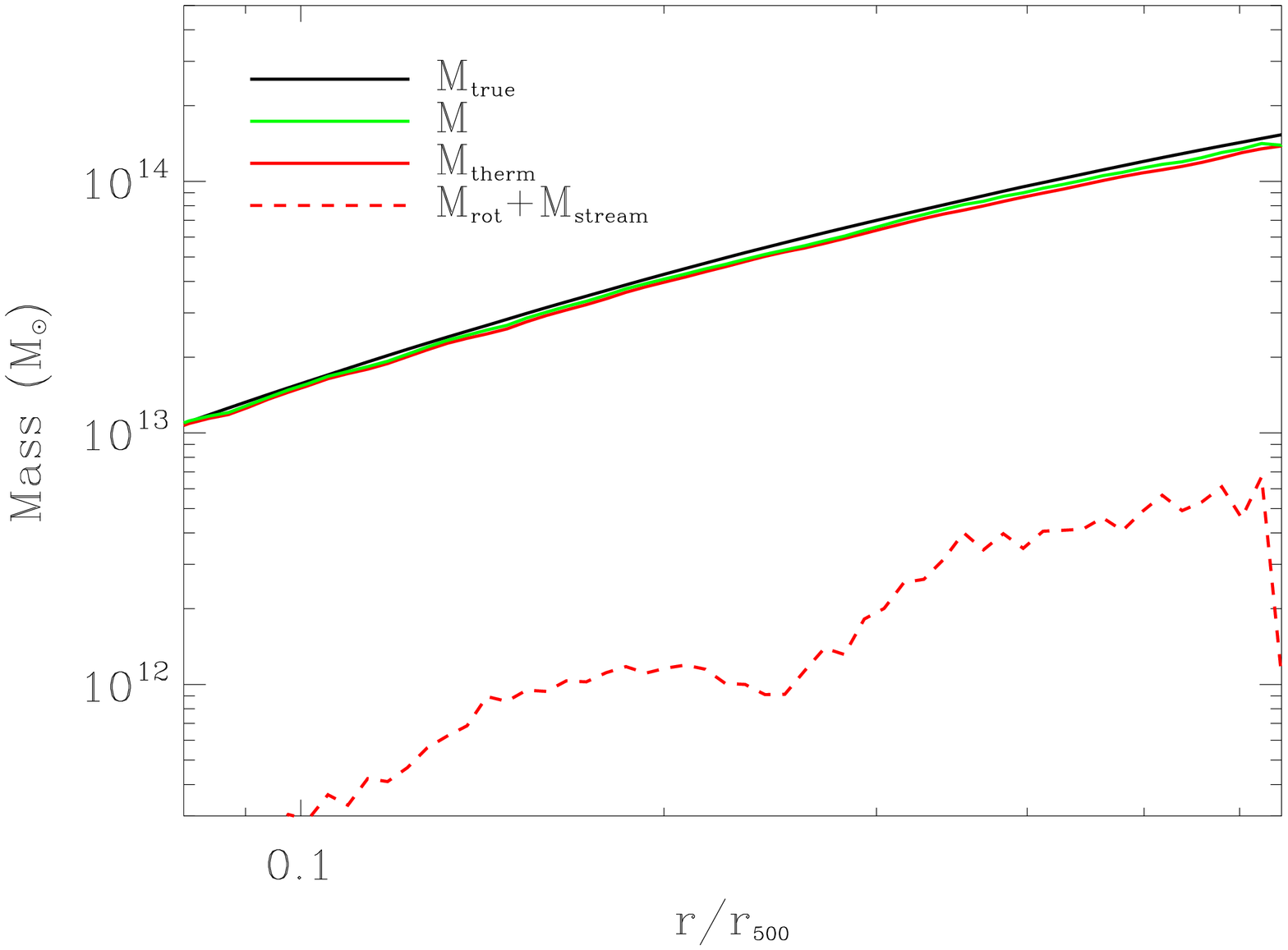}
\includegraphics[angle=90,scale=0.35]{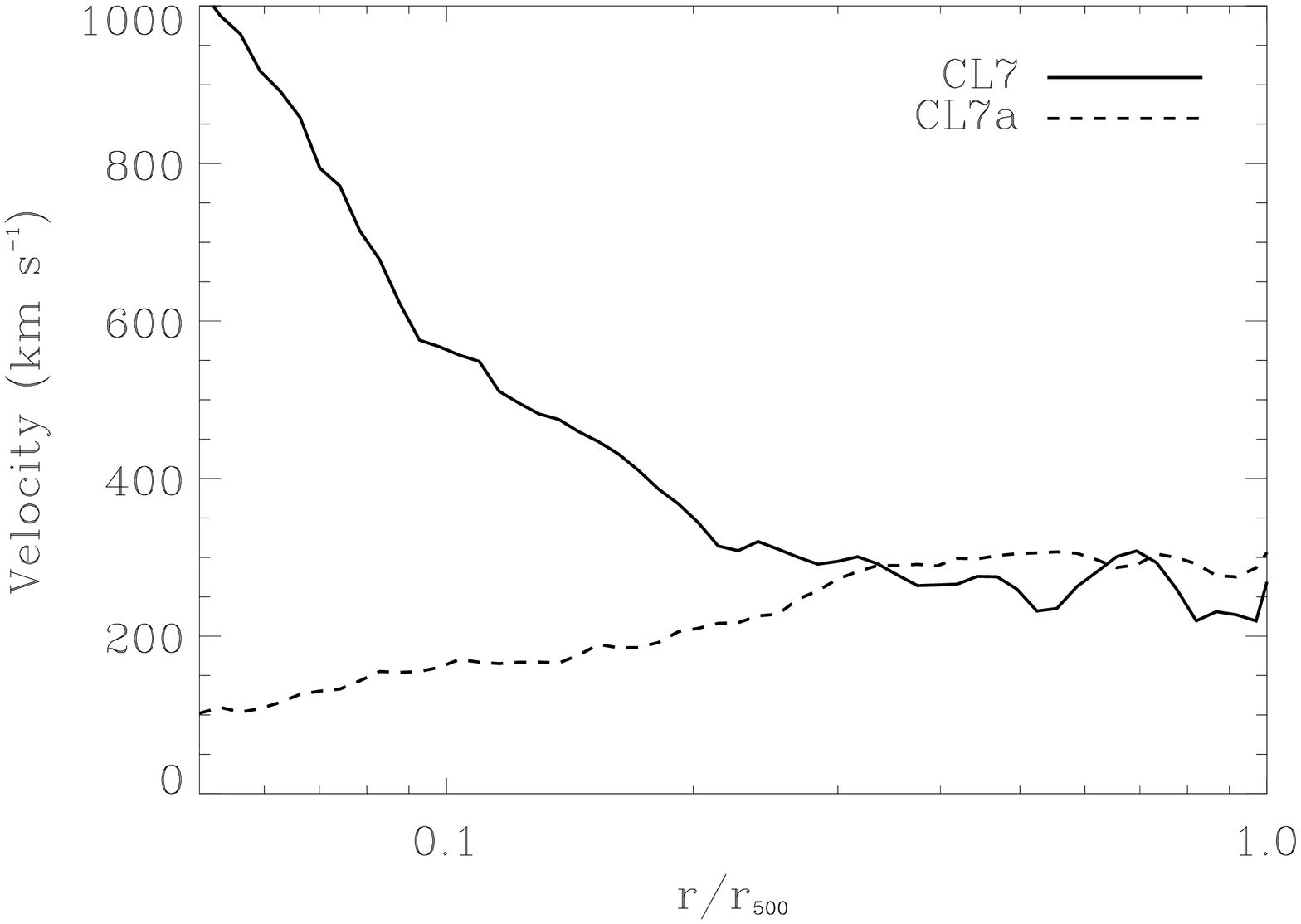}
\end{center}
\caption{Left panel: same as the left panel of Figure \ref{m7} but for
  CL7a. Right panel: rotational velocity as a function of radius for CL7 (solid line) and CL7a (dashed line).}
\label{cl7a}
\end{figure*}

In the left panel of Figure~\ref{cl7a} we plot mass profiles
for various terms in CL7a, applying the same method that we used for
CL7. For this simulation unfortunately part of the temperature data is
corrupted, and we have only those within $1h^{-1}$ Mpc box and compute
the mass profiles out to $\sim 0.6\ r_{500}$. Clearly, for the simulation without radiative cooling,
$M_{therm}$ closely follows the total mass, and the contribution from
$\left(M_{rot}+M_{stream}\right)$ is less than 5\% over most
radii. This is in sharp contrast to the case of CL7, in which the
contribution from the $\left(M_{rot}+M_{stream}\right)$ can be up to
10 -- 20\%, especially in the inner regions (see left panel of
Figure~\ref{m7}). This suggests that rotation and streaming motion
contribute less to the gas dynamics in the NC simulations; i.e., that
without radiative cooling.

To illustrate the different between the CSF and NC simulations, we
also plot the rotational velocities in the CL7 and CL7a. The right
panel of Figure~\ref{cl7a} shows the rotational velocity as a function
of radius. We take the equatorial plane and average the gas velocity
projection in the plane in each radial bin. We notice that both NC and
CSF simulations show similar gas velocities at $\gtrsim 0.3 r_{500}$,
between 200 and 300 $\rm km\ s^{-1}$. However, at the inner regions,
the gas volocity in the CSF simulation rises sharply to above 1000
$\rm km\ s^{-1}$, while the NC simulation shows a steady decline.

The substantial gas rotation has a strong impact on the gas dynamics
in the central regions. This can be quantified by the ratio ($R_{E}$)
between the kenetic energy in rotation ($KE_{rot}$) and the thermal
energy ($E_{therm}$). For the CSF cluster CL7, we have 
 \begin{eqnarray}
R_{E} & \equiv & \frac{KE_{rot}}{KE_{therm}} = \frac{(1/2)\rho_g
  v^2_{rot}}{(3/2)n_g k_B T_g} \\
 & = & 0.23 (v_{rot}/600\ km\ s^{-1})^2(T_g/3.5\ keV)^{-1},
\end{eqnarray}
where $v_{rot}$ is the rotation velocity. Taking the typical rotational
velocity of $\sim 600\rm\ km\ s^{-1}$ at $\sim 0.1r_{500}$ in CL7 (see
Figure~\ref{cl7a}), we estimate the rotational kinetic energy can be as high as about 23\% of
the thermal energy. This is actually a conservative lower limit since
the gas temperature peaks at $\sim 3.5$ keV (Nagai et al.~2007). It
can be as low as $\sim 1$ keV in the innermost regions. This ratio
decreases to about 10\% between 0.2 and 0.3 $r_{500}$. However, for
CL7a, due to the relative low velocity, this ratio stays below 10\%
at most radii.In fact, in the adiabatic simulation the $R_E$ value at
$0.1\ r_{500}$ is only
about $1/10$ the value obtained for the simulation with cooling and star
formation.   

\section{ICM Rotation and Ellipticity}

\subsection{Motivation}

The substantial ICM rotation predicted by the simulations for relaxed
CDM clusters needs to be tested by observations. The straightforward
approach to detect rotation and other bulk motion of the ICM in
clusters is to measure Doppler shifts and broadening of emission lines
in their X-ray spectra. For a typical bulk-motion velocity of $\sim
500\rm\ km\ s^{-1}$, Sunyaev, Norman \& Bryan~(2003) estimated that
the Doppler shift of an iron line at 6.7 keV would be $\sim 10$ eV.
The X-ray Spectrometer (XRS) on Astro-E2 would have been the first
instrument to perform such measurements for many clusters.  But its
failure soon after launch means we must await the next generation of
X-ray satellites, such as {\sl Constellation-X}, {\sl XEUS} and {\sl Spectrum-X}, for
direct measurements of ICM motions in many clusters.

In the classic cooling flow model, gas rotation is expected near the
center of the flow due to mass and angular momentum conservation
(e.g., Mathews \& Brighenti~2003). In the case of elliptical galaxies,
the X-ray images are expected to be considerably flattened toward the
equatorial plane (e.g., Brighenti \& Mathews~1996;  Mathews \&
Brighenti~2003). X-ray morphological information, such as ellipticity,
can be used to probe this process then. X-ray studies that assume
negligible ICM rotation indicate that both the radial profile (e.g.,
Pratt \& Arnaud 2003; Lewis, Buote, \& Stocke 2003; Vikhlinin et al.\
2006; Gastaldello et al.\ 2007) and ellipticity (Buote \& Canizares
1996) of the gravitating mass of nearby clusters agree with the radial
profile and ellipticity of the dark matter of simulated CDM
clusters. Therefore, if the ICM of the simulated CDM clusters in our
study is flattened substantially by rotation, their X-ray images on average
should have larger ellipticities than observed clusters.  

Since X-ray ellipticity can be measured precisely for many clusters,
below we compare results for the simulated CDM clusters to a sample of
observed clusters. We will also compare to a simulated CDM clusters
that involves no radiative cooling and star formation. This comparison
will help us understand the role of over-cooling. We note that previous studies found that the
simulation we use here reproduces reasonably well the spherically
averaged global ICM properties of observed clusters (Nagai et
al.~2007; Kravtsov et al.~2006).

\subsection{Ellipticity of Relaxed CDM Clusters} \label{ellsim}

\begin{figure}[t]
%\centering
\centerline{\includegraphics[angle=90,scale=0.4]{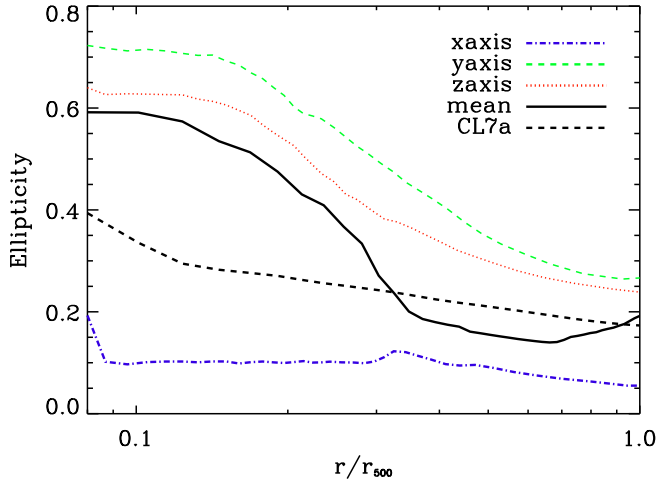}}
\caption{X-ray ellipticity profiles of the simulated cluster CL7
obtained by projecting the cluster along the $x$-axis (blue), $y$-axis
(green), and $z$-axis (red). The dark solid line is the mean profile
obtained by averaging the profiles from 100 projections of random
orientation (see text), and the dark dashed line is the mean profile
of the cluster in the simulation without radiative cooling and star
formation.} 
\label{ell}
\end{figure}

We begin by computing the ellipticity profiles of the X-ray images of
the relaxed simulated cluster CL7 for the three orthogonal projections
shown in Figure \ref{image}. The iterative moment-based procedure we
use to evaluate the ellipticities of the simulated X-ray images is the
same as that discussed below for the observations of real clusters (\S
\ref{obsell}).  We show the X-ray ellipticity profiles for CL7 in
Figure \ref{ell}.

The X-ray ellipticities of the $x$-projection are small and have a nearly constant value ($\approx 0.10$) from $0.1-0.7\, r_{500}$
eventually falling to ($\approx 0.05$) at $r_{500}$. The very round
X-ray isophotes are expected in this projection since one is
essentially looking down the axis of rotation. In contrast, the
ellipticities are much larger and show a much stronger variation with
radius in the other two projections. For the $y$-projection, the
ellipticity is $0.6-0.7$ between radii $0.1-0.2\, r_{500}$, declines
to $\approx 0.5$ at $0.3\, r_{500}$, and continues to fall with
radius until reaching a value of $\approx 0.25$ at $r_{500}$. The
$z$-projection has ellipticities similar to the $y$-projection but
with $\approx 10\%$ smaller values for most radii. 

The X-ray ellipticities for radii $0.1-0.3\, r_{500}$ in the $y$ and
$z$ projections are considerably larger than those of the dark matter
($\approx 0.4$), which also decline with increasing radius.  The
larger X-ray ellipticities with respect to the dark matter imply
substantial rotational flattening of ICM, because (1) the
gravitational potential of an ellipsoid is never flatter than the
mass distribution that generates it, (2) for (non-rotating)
single-phase hot gas in hydrostatic equilibrium, the X-ray emissivity
has the same shape as the gravitational potential, independent of the
temperature profile of the gas (``X-Ray Shape Theorem'', Buote \&
Canizares 1994, 1998), and (3) $M_{rot}$ dominates $M_{stream}$ and
$M_{turb}$. 

As a consistency check, we note that at 0.1 $r_{500}$ both $M_{rot}/M_{therm}$
and $R_E$ (section 2.4) are about 0.25 for CL7, indicating substantial
rotational support. Moreover, $1+R_E \sim 1.25$ provides an approximate
indication of the axial ratio induced by rotation, and this can explain
most of the excess X-ray ellipticity over that produced by the flattened
dark matter potential.

To make a proper comparison of the X-ray ellipticity profile of the
simulated CDM clusters to the average ellipticity profile of observed
clusters we average the ellipticity profiles of CL7 obtained by
viewing 100 projections of random orientation $(\theta,\phi)$.  The
average ellipticity profile is calculated by averaging all the
ellipticity values of a particular radius obtained for the 100
simulations. To achieve fast computation, we approximate the X-ray
emissivity in each cell as $\propto n_g^2 T^{1/2}$, appropriate for
thermal bremsstrahlung radiation.  We verified this approximation by
comparing to the profiles for the three orthogonal projections
projections computed using the full expression for the plasma
emissivity.

The angle-averaged X-ray ellipticity profile for CL7 is shown in
Figure \ref{ell}. The average profile is similar to that from the $y$
and $z$ projections but with somewhat smaller values; i.e., the
ellipticity declines from $\approx 0.6$ at $0.1\, r_{500}$ to $\approx
0.3$ at $0.3\, r_{500}$ and then settles to a value between 0.15-0.20
at larger radii.

Using the same procedure, we computed average ellipticity profiles for
the other ``relaxed'' clusters in the simulation. Since below we
consider observations of clusters (rather than lower mass groups), we
only include the relaxed simulated clusters with masses above
$10^{14}h^{-1}M_{\odot}$; i.e., CL3, CL5, CL7, and Cl104.  We
angle-averaged the profiles of these systems and plot the combined
average profile of these four systems in Figure~\ref{rosat}. The
combined profile is very similar to that of CL7 (Figure
\ref{ell}).

\subsection{Comparison to Adiabatic Simulation}

For comparison, in Figure\ref{ell} (dark dashed line) we show the average
ellipticity profile of CL7a obtained from the simulation without cooling
or star formation (NC). Since we don't have the temperature data outside of
the $1h^{-1}$ Mpc box and the X-ray emissivity has a very weak
dependency on temperature, we approximate the X-ray emissivity in each
cell as $\propto n_g^2$. We test our results within the $1h^{-1}$ Mpc
box and do not find any major difference between these two
approximations. Within the central regions $(\sim 0.25 r_{500})$ the
ellipticity of the NC cluster is considerably smaller compared to the
CSF simulation. In fact, the ellipticity profile remains nearly constant
with a value $\sim 0.25$ for radii larger than $0.1 r_{500}$. The much smaller
ellipticity in CL7a is consistent with the much smaller rotational
support compared to CL7. From Figure 6 one can see that the very
different gas velocities near $0.1 r_{500}$ translate to $R_E$ values in CL7a
a factor of 10 smaller than in CL7. At $0.3 r_{500}$, the gas velocities of
CL7 and CL7a are similar, as are their average X-ray ellipticities. The
slightly smaller ellipticity values of CL7 at larger radii reflect both
the slightly smaller gas velocities over that range (see Figure 6) and
the fact that CL7 is slightly more centrally concentrated than CL7a.
Clearly, the gas dynamics and morphology differ substantially,
particularly within $\sim 0.3 r_{500}$, depending on whether cooling is included
in the simulations.

\subsection{Ellipticity of Observed Clusters}

\begin{figure*}[t]
\centering
\includegraphics[angle=90,scale=0.6]{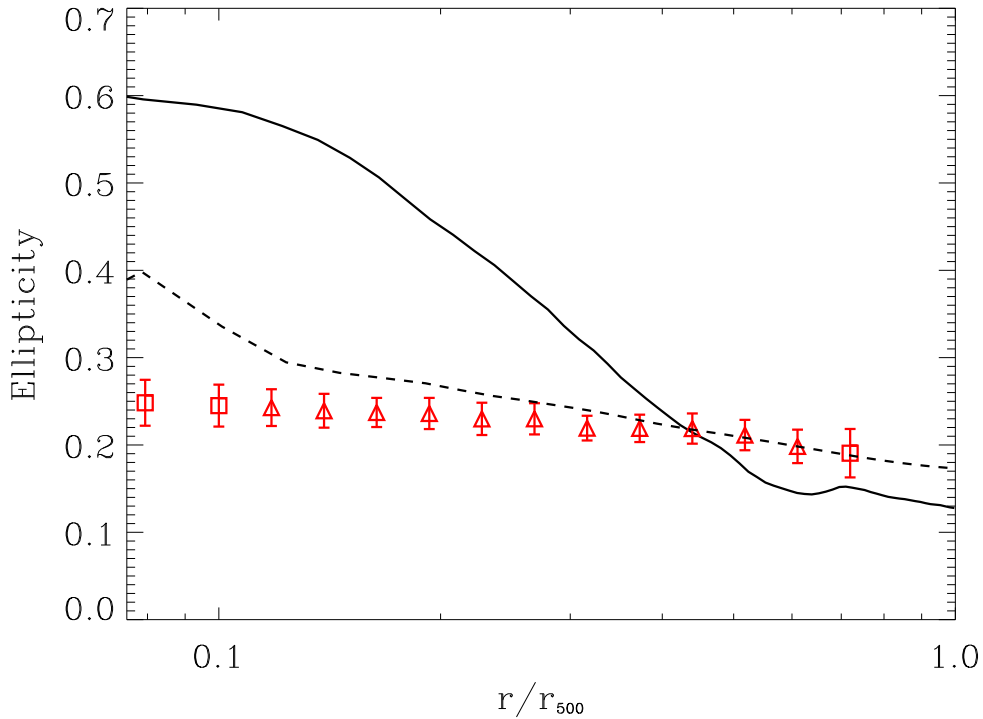}
\caption{Comparison of the average X-ray ellipticity profiles of the
``relaxed'' simulated CDM clusters (solid black) and the observed
clusters (dashed red). For the observed clusters, the inner parts of
the ellipticity profile are based on the \chandra\ data, while the
outer parts are derived from \rosat\ PSPC. The error-bars on the 
observed profile represent the uncertainty on the mean arising from
the intrinsic scatter in the measured data. Triangles denote data-points
where all 9 observed clusters overlapped, and squares indicate the 
regions where fewer clusters were used to compute the average profile.
Note that none of the
observed clusters have ellipticities in excess of $\sim$0.3 and none
show evidence of profiles which steeply rise towards the
center. Dashed line is the ellipticity profile of CL7a from the simulation without radiative cooling or star
formation.
}
\label{rosat}
\end{figure*}

\subsubsection{Sample definition}

In order to effect a comparison between the observable properties of the simulated clusters and real
data, we selected a sample of nearby {\em relaxed} clusters observed with the \rosat\ PSPC and with 
\chandra. We opted to use these complementary satellites since the
large, unobstructed field of view  ($\sim$20\arcmin\ radius) makes \rosat\ ideal for 
the unbiased computation of ellipticities 
out to radii in excess of $\sim$500~kpc for most interesting systems. Its
modest spatial resolution (0.5\arcmin\ FWHM at 1~keV; corresponding to
$\sim$50~kpc at $z=$0.1) of the PSPC will not significantly bias the ellipticity calculation within
radii \gtsim 100--200~kpc. In contrast \chandra's excellent 
spatial resolution ($\sim$0.5\arcsec, corresponding to scales of \ltsim 2~kpc even at redshifts as 
high as $\sim$0.2), but modest field of view (\ltsim 4\arcmin\ radius on any single CCD) 
allows ellipticities to be measured from scales ranging from $\sim$1 to $\sim$100~kpc.
We did not use the \rosat\ HRI due to its smaller field of view and lower sensitivity than the PSPC.

To define a sample of relaxed clusters we selected the 10 clusters from the PSPC sample of Buote \& Tsai~(1996)
which had the lowest \pthree\ ratio within an aperture of 500$h_{80}^{-1}$~kpc. We adopted the \pthree\ ratio
since it is sensitive to unequal-sized mergers (Buote \& Tsai~1995), and therefore allows us to eliminate
obviously disturbed objects. Although there is evidence of correlations between \pthree\ 
and \ptwo\ (which is sensitive to the cluster ellipticity) {\em for disturbed systems}
(Buote \& Tsai~1996), we would expect a relaxed cluster to have \pthree $\simeq$0, and so
this selection should not bias us towards low-ellipticity (relaxed) clusters.
We excluded one cluster, A\thin 2255, since it has a very extended surface brightness profile that implies 
that it is not in a relaxed state. We also excluded A\thin 1651 since the existing \chandra\ observation
is very shallow and the cluster is centered close to a chip gap (restricting the range of radii over
which the ellipticity can be computed). Although it is difficult to choose relaxed clusters which 
exactly match the mass distribution of the simulated clusters, to increase the coverage
of lower-mass systems we included A\thin 2589, which just missed our \pthree\ cut but is nonetheless
very relaxed and has a Virial mass $\sim 3\times 10^{14}$\msun\
(Zappacosta et al.~2006).
The selected objects, along with properties from Buote \& Tsai~(1996) are shown in
Table~\ref{table_sample}. To ensure a fair comparison, we also examine
the \pthree\ ratio for the simulated clusters, and find they are in
range of $10^{-8} - 10^{-10}$, consistent with those of observations.

\subsubsection{\chandra\ data reduction}

Archival observations of the clusters were drawn from the public \chandra\ archive, 
and the data were reduced and analyzed with the 
\ciao\ 3.4 and \heasoft\ 5.3.1 software
suites, in conjunction with \chandra\ calibration database (\caldb)
version 3.3.0.1. To ensure up-to-date calibration, all data were
reprocessed from the ``level 1'' events files, following the standard
\chandra\ data-reduction threads\footnote{{http://cxc.harvard.edu/ciao/threads/index.html}}.
We applied the standard correction to take account of the time-dependent gain-drift
and, where possible, charge transfer inefficiency, 
as implemented in the \ciao\ tools. To identify periods of enhanced
background (``flaring''), which seriously degrades the signal-to-noise (S/N)
we accumulated background light curves for each data set from
low surface-brightness regions of the active chips, excluding obvious 
point-sources. Periods of flaring were identified by eye and
excised. The final exposure times are listed in Table~\ref{table_sample}. For each
data set we generated a full resolution image in the energy 0.3--7.0~keV and corresponding
exposure map, computed at an energy of 1.7~keV. Based on experience, such a monochromatic 
exposure-map is sufficient for our present purposes. 

Since our adopted procedure to calculate the ellipticity of a cluster involves taking moments of this 
image,
the presence of bright point-sources (most likely AGN) can lead to errors in the computation. 
It is therefore essential to identify and remove these objects from the image.
Point-sources were detected by application of the CIAO task {\tt wavdetect}, which was set to search for structure at scales of 1, 2, 4, 8 and 16 
pixels, and supplied with the exposure-maps to minimize spurious detections at the image boundaries.
The detection threshold was set to $10^{-6}$, corresponding to \ltsim 1 spurious source
detections per cluster. All detected sources were confirmed by visual inspection, and, for each,
appropriate elliptical regions containing approximately 99\%\ of its photons were generated.
The detection algorithm also generated a ``normalized background'' image, which is effectively a 
smoothed, flat-fielded image, having removed the point-sources found by {\tt wavdetect}.

To generate a cleaned image, suitable for the calculation of ellipticities, 
the data within each of the elliptical point-source regions were 
replaced with Poisson-deviated noise.
For each pixel, the mean of the Poisson deviate was drawn from
the corresponding pixel on the normalized background image, taking into account variations in the exposure-map. 
Provided the detection algorithm accurately characterizes all the point-sources this is a very robust method
for source removal. However, {\tt wavdetect} occasionally mis-identifies extended cluster
emission as a point-source, or mis-estimates the radius enclosing most of the source photons, 
and so the background map may be locally incorrect. 
The point-source free, ``cleaned'', image was, therefore, visually inspected to assess how accurately
sources were removed. In cases where the source replacement was obviously imperfect, suspicious sources were
replaced by assuming the mean (flat-fielded) count-rate per pixel is uniform over the point-source region
(this is most correct if the source is at a large radius from the cluster center). This
count-rate was estimated for each source by centering a circular extraction region on the source centroid and 
expanding its radius iteratively until at least 50 photons were found within it, taking care to exclude
all photons within any of the source regions. Since the exposure may vary over the pixels within a given
source region (especially near the edge of the image), the exposure-map was used to correct this rate
on a pixel-by-pixel basis before Poisson noise was added.
The final ``cleaned'' images were
visually inspected to confirm reliable source removal.  Provided the brightest of the point-sources in an image 
are removed in this way, any residual (unidentified) sources should not significantly bias our 
results (Buote \& Tsai~1995).  Finally, for the computation of the ellipticity profiles, we flattened the 
image by dividing it by its exposure-map. To confirm the accuracy of the exposure correction, we 
visually inspected each image (having first mildly smoothed it with a $\sim 10$\arcsec\ Gaussian filter
to aid the eye) for obvious signs of poor flat-fielding. In general, away from the gaps between CCDs, the exposure 
correction appeared reliable.

\begin{deluxetable*}{llllllllll}
\tablewidth{0pt}
\tabletypesize{\footnotesize}
%\rotate
\tablecaption{The cluster sample\label{table_sample}}
\tablehead{
\colhead{Cluster }&\colhead{Redshift }&\colhead{\rfive}&\colhead{\ptwo }&\colhead{\pthree }&\multicolumn{3}{c}{\chandra}&\multicolumn{2}{c}{\rosat} \\
\colhead{}&\colhead{ }&\colhead{(kpc)}&\colhead{}&\colhead{ }&\colhead{ObsID }&\colhead{Exposure }&\colhead{$R_{max}$ }&\colhead{ObsID }&\colhead{Exposure } \\
\colhead{ }&\colhead{ }&\colhead{}&\colhead{ }&\colhead{ }&\colhead{
}&\colhead{(ks) }&\colhead{(kpc)}&\colhead{ }&\colhead{(ks) }}
\startdata
A2589 &$0.041$ &$820^e$ &$(2.6^{+0.8}_{-1.0}) \times 10^{-6}$ &$(1.1^{+5.9}_{-0.9}) \times 10^{-8}$ &7190 &$53$ &$250$ &800526 &$5.9$ \\
A2597 &$0.085$ &$900^b$ &$(1.0^{+0.3}_{-0.4}) \times 10^{-6}$ &$(1.6^{+31.}_{-1.3}) \times 10^{-9}$ &7329 &$58$ &$320$ &800112 &$6.2$ \\
A2199 &$0.030$ &$990^c$ &$(5.9  \pm 1.00) \times 10^{-7}$ &$(5.1^{+12.}_{-3.5}) \times 10^{-9}$ &497 &$18$ &$310$ &150083,800644 &$43.$ \\
MKW3s &$0.043$ &$1200^a$ &$(1.6^{+0.5}_{-0.6}) \times 10^{-6}$ &$(5.3^{+33.}_{-4.4}) \times 10^{-9}$ &900 & 57 &230 &800128 &$7.5$ \\
A1795 &$0.062$ &$1240^d$ &$(1.3  \pm 0.1) \times 10^{-6}$ &$(3.0^{+22.}_{-3.0}) \times 10^{-10}$ &5287 &$14$ &$310$ &800055,800105 &$54.$ \\
A2052 &$0.035$ &$1290^a$ &$(1.3  \pm 0.4) \times 10^{-6}$ &$(3.3^{+26.}_{-2.7}) \times 10^{-9}$ &5807 &$130$ &$290$ &800275(a01,n00) &$6.7$ \\
A1413 &$0.14$ &$1300^d$ &$(5.2^{+2.3}_{-1.7}) \times 10^{-6}$ &$(2.2^{+110}_{-1.3}) \times 10^{-9}$ &5003 &$75$ &$270$ &800183 &$6.4$ \\
A2029 &$0.077$ &$1360^d$ &$(1.4  \pm 0.3) \times 10^{-6}$ &$(3.1^{+6.7}_{-2.9}) \times 10^{-9}$ &4977 &$77$ &$300$ &800161,800249 &$12.$ \\
HydraA &$0.052$ &$1390^a$ &$(4.9  \pm 1.3) \times 10^{-7}$ &$(4.4^{+10.0}_{-3.6}) \times 10^{-9}$ &4969 &$86$ &$260$ &800318 &$16.$ 
\enddata
\tablecomments{The sample of ``relaxed'' galaxy clusters, sorted in order of increasing \pthree\
ratio, as obtained from Buote \& Tsai~(1995). Also shown are the cluster redshift, the mean cluster
temperature (kT), the \ptwo\ and \pthree\ ratios from Buote \& Tsai~(1996), the observation numbers (``ObsID'') of each \rosat\ PSPC and \chandra\ data set used in the 
present paper, and the total cleaned exposure time, following background screening (``Exposure''). 
For the \chandra\ data, we also indicate the maximum radius to which the data was used in computing the 
ellipticity profile ($R_{max}$)\\
\rfive\ taken from: $^a$Finoguenov et al.\ (2001), $^b$Pointecouteau et al.\ (2005),
$^c$Sanderson et al.\ (2006), $^d$Vikhlinin et al.\ (2006), $^e$Zappacosta et al.\ (2006) 
}
\end{deluxetable*}

\subsubsection{\rosat\ data reduction}

We obtained RDF data for each cluster from the \heasarc\ archive, which were 
reduced and processed with \heasoft\ 5.3.1 and CIAO 3.3.0.1. 
The data were initially screened to remove 
(probably spurious) bursts of events with the {\tt burst} task. To ensure conservative screening criteria,
we generated a PSPC ``makefilter file'' using the {\tt pcfilt} task, and screened out events 
corresponding to a master veto rate in excess of 170 $s^{-1}$.
A light curve was generated from outer parts of the field of view, chosen to avoid obvious bright 
point-like or extended sources, and was examined by eye to identify periods of enhanced background.
Any data taken during such ``flares'' were excluded from further analysis. Full field of view images 
were generated in the 0.4--2.0~keV band, with a pixel-size of 15\arcsec, 
and corresponding exposure maps were generated with the 
{\tt pcexpmap} task. In addition, a circular cut-out region, centered on the peak of the cluster emission
and extending as far as the shadow of the inner PSPC support ring, was created using the {\tt dmcopy} task
for both the image and exposure-map. Such a cut prevents the support structure shadows (which are not
perfectly removed during flat-fielding) from compromising the calculation of the ellipticities.
For those objects with multiple pointings (Table~\ref{table_sample}),
the images and exposure-maps were summed to produce a single ``merged'' image and exposure-map, 
using custom software. Before addition, each image was aligned with the image having the longest 
exposure by shifting it an integer number of  pixels in the x and y directions.

Point-sources were detected in the merged cut-out image with the  {\tt wavdetect} algorithm, 
which was set to search for structure at scales of 1, 2, 4 and 8 pixels,
and supplied with the exposure-maps to minimize spurious detections at the image boundaries.
The detection threshold was set to $10^{-6}$, corresponding to \ltsim 0.1 spurious source
detections per image. After visual confirmation of the detected sources, the images were ``cleaned''
of point-sources and flat-fielded, in exactly the same manner as  the \chandra\ data.
To prevent over-correction during flat-fielding (particularly in areas where the 
exposure-correction may be questionable), we reset to zero any pixel for which the exposure map fell below 30\%
of its peak value, and reset the maximum radius inside which the ellipticity can be computed so as not to contain
any such pixels. 

\subsubsection{Ellipticity Profiles} \label{obsell}

To compute the ellipticity profiles, we adopted a moment-based method
that was first developed by Carter \& Metcalfe (1980) and later was
implemented in the study of the {\sl ROSAT} and {\sl Chandra} image of NGC
720 (see, e.g., Buote \& Canizares 1994; Buote et al.~2002) and a
sample of five {\sl ROSAT} clusters (Buote \& Canizares~1996). In this
method, the two-dimensional principal moments of inertia are computed within an
elliptical region. The ellipticity, $\epsilon$, is defined as the
square root of the ratio of the principal moments. We refer the
readers to Buote \& Canizares (1994) and Buote et al.~(2002) for
details. 

For each cluster we visually inspected the \chandra\ image for evidence of disturbances in the core,
such as bubbles, shocks or cavities, which are most likely a
consequence of AGN activity (B{\^i}rzan et al.~2004; Forman et
al.~2005; Fabian et al.~2006). Such effects will distort the ellipticity
profile computed within the disturbed region but, crucially, the physics of these interactions is not captured
by the simulations. Therefore, for each cluster, we estimated the extent of the disturbed region and
ignored the ellipticity profile computed within this radius. The moment method we adopted to compute the
profile does incorporate information from the pixels in the suspect region. However,
the ellipticity is strongly biased towards pixels at large radius and so the method should still yield
reliable results. The \chandra\ profiles were computed to as large a radius as possible before 
a circular aperture of that radius would come in contact with a chip gap or, alternatively, the 
error-bars on the profile became large. The maximum extent of each  \chandra\ profile is given
in Table~\ref{table_sample}. Outside this radius, the ellipticity data-points were taken from the
\rosat\ data, for which the profile was computed until the edge of the ``cut-out'' image. 
In general, we found that the \rosat\ and \chandra\ profiles matched up well.

In Table~\ref{table_ellipticity} we list the observed ellipticity profiles
derived for each of the clusters in our sample.
 In general, the profiles are relatively flat, with a peak ellipticity of $\sim$0.2--0.3. We do
not find any evidence of sharply rising profiles.

\subsection{Comparison of Observed and Simulated CDM Clusters}

We show the average X-ray ellipticity profiles of the observed
clusters and the simulated CDM clusters ($M>10^{14}h^{-1}M_{\odot}$)
in Fig~\ref{rosat}, with $1\sigma$ statistical errors on the
mean observed values. We also list the average observed ellipticities
and their errors in the last column of Table~\ref{table_ellipticity}. The differences between the profiles are very
striking.  While the observed clusters have a nearly constant
ellipticity profile ($\sim 0.15$), the simulated CSF clusters have much
larger ellipticity values ($\sim 0.3-0.6$) at small radii ($0.1-0.4\,
r_{500}$) before declining to $\sim 0.15$ near $r_{500}$. For radii
above $\approx 0.6r_{500}$ both the shape and normalization of the
profiles are similar for both the simulated and observed
clusters. However, the ellipticity profile of the NC cluster CL7a shows much
better agreement with the observed clusters over all radii examined.
The relatively small deviations of this single cluster (formed in the
simulation without cooling or star formation) from the average
ellipticity profile of the observed clusters is consistent with the
stochastic variation between the observed clusters (see Table A2).

The pronounced discrepancy over $0.1-0.4\, r_{500}$, taken together
with the discussion in \S \ref{ellsim}, demonstrates clearly that
the ICM in the simulated clusters rotates much more rapidly than in
the observed clusters. Furthermore, this excess ICM rotation in the
simulated CDM clusters extends out to $\approx 0.6r_{500}$ since that
is where the ellipticity profiles of the observed and simulated
clusters diverge. The averaged profiles for the CSF clusters also are
slightly lower than the observations at $>0.6\ r_{500}$, indicating either that the CSF
clusters may rotate a little slower than the observed clusters at larger
radii and, perhaps more likely, the CSF clusters are slightly more
centrally concentrated.

\section{Discussion and Conclusion}

Over the past decade theoretical studies of cosmological
hydrodynamical simulations have emphasized the important role of
random turbulent motions in providing pressure support of the ICM in
galaxy clusters. Using the set of high-resolution CDM simulations of
Nagai et al.\ (2007) we have shown instead that rotational support of
the cluster weight exceeds that of random turbulence from $0.1-0.5\,
r_{500}$ and remains comparable out to $\approx 0.8r_{500}$ for the
clusters classified as the most relaxed in the simulations.

When we compared the average ellipticity profile of the
relaxed CDM clusters to that obtained for a set of 9 real nearby
clusters observed by {\sl Chandra} and {\sl ROSAT}, we found the
simulated CDM clusters are significantly flatter within $\approx
0.4r_{500}$ and the profile shapes differ interior to $\approx
0.6r_{500}$. By comparing simulations with and without radiative
cooling and star formation, we conclude that this flatness is mainly
caused by gas over-cooling and large-scale rotation. 

The substantial ICM rotation present in the relaxed simulated CDM
clusters very likely indicates that a classical ``rotating
cooling flow'' (Nulsen, Stewart, \& Fabian 1984; Kley \& Mathews 1994;
Brighenti \& Mathews 1996) operates in those systems, implying large
mass deposition rates which are not observed.  At very small radii
($<0.1 r_{500}$) it is well-known that cosmological hydrodynamical
simulations fail to reproduce observed ICM properties because of
over-cooling, but since the ICM rotation we have discussed extends to
much larger radii, the over-cooling problem apparently does as well.

We stress that the Nagai et al.\ (2007) simulations are not unusual in
regards to displaying signatures of ICM rotation. Recently, Jeltema et
al.\ (2008) compare X-ray images of simulated and observed clusters
using power ratios (Buote \& Tsai 1995). They find that $P_2/P_0$,
which is similar to ellipticity, computed within a 0.5~Mpc aperture
for their relaxed simulated CDM clusters is slightly larger than for
the observed clusters. They also find that distribution of $P_3/P_0$,
which is sensitive to asymmetry but not ellipticity, is consistent for
the simulated and observed clusters, implying that it is very unlikely
that the $P_2/P_0$ (and hence ellipticity) discrepancy can be
attributed to the chosen cosmology.  Since 0.5~Mpc corresponds
roughly to $0.5r_{500}$ for clusters, the relatively weak discrepancy
is consistent with our results (see Figure \ref{rosat}).

Because the unphysical ICM rotation in the relaxed CDM clusters
extends out to a large radius ($\approx 0.6r_{500}$), ICM quantities
determined within these regions in the simulations need to be
reconsidered. For example, it has been noted by several authors that
X-ray measurements of $M(<r_{500})$ for relaxed clusters assuming
hydrostatic equilibrium underestimate the actual mass by $5-20\, \%$
(e.g., Rasia et al.\ 2006; Nagai et al.\ 2007; Burns et al.~2007). But since the X-ray
analysis involves fitting gas density and temperature profiles of the
ICM over a large radial range interior to $r_{500}$, the region of
unphysical ICM rotation is included in the analysis. (Note that
excluding large radial regions can strongly bias X-ray mass estimates
-- Gastaldello et al.\ 2007.) While we do not expect these X-ray mass
underestimates to change radically for relaxed simulated CDM clusters
without unphysical ICM rotation, given that $M_{rot}\sim M_{turb}$ out
to $\approx 0.8r_{500}$ in the simulated clusters we have studied, it
is not unreasonable to expect the mass underestimate to be cut in
half (see Section 2.4). This would be of considerable importance to studies using X-ray
determinations of cluster masses for precision cosmology.

{\it Acknowledgments:} We thank Daisuke Nagai for providing the sample
of simulated clusters and a careful reading of our manuscript. We also
thank the referee for helpful comments. We thank Renyue Cen, Fabio Gastaldello, Andrey Kravtsov,
Luca Zappacosta for helpful discussions. We gratefully acknowledge
partial support from NASA through {\sl Chandra} Award Number
GO6-7118X, GO6-7120X, GO6-7671X, and NNG04GE76G, issued through the
Office of Space Sciences Long-Term Space Astrophysics Program.

\appendix
\section{Observed cluster ellipticity profiles}
We show in Table~\ref{table_ellipticity} the observed ellipticity profiles of each of 
the clusters considered in \S~\ref{obsell}.

\clearpage

\LongTables
\begin{landscape}
\begin{deluxetable}{lllllllllll}
\tablewidth{0pt}
%\rotate
\tablecaption{Observed cluster ellipticity profiles \label{table_ellipticity}}
\tabletypesize{\scriptsize}
\tablehead{
\colhead{r/\rfive} & \multicolumn{10}{c}{Observed cluster}\\
\colhead{ }&\colhead{A2589 }&\colhead{A2597 }&\colhead{A2199 }&\colhead{MKW3s }&\colhead{A1795 }&\colhead{A2052 }&\colhead{A1413 }&\colhead{A2029 }&\colhead{HydraA }&\colhead{Average} \\
}
\startdata
$0.010$ &\ldots &\ldots &\ldots &\ldots &\ldots &\ldots &\ldots &$0.090\pm 0.031$ &\ldots &\ldots\\
$0.013$ &\ldots &\ldots &\ldots &\ldots &\ldots &\ldots &\ldots &$0.14\pm 0.03$ &\ldots &\ldots\\
$0.016$ &\ldots &\ldots &\ldots &\ldots &\ldots &\ldots &\ldots &$0.16\pm 0.02$ &\ldots &\ldots\\
$0.020$ &\ldots &\ldots &\ldots &\ldots &\ldots &\ldots &\ldots &$0.20\pm 0.01$ &\ldots &\ldots\\
$0.025$ &\ldots &\ldots &\ldots &\ldots &\ldots &\ldots &\ldots &$0.198\pm 0.010$ &\ldots &\ldots\\
$0.032$ &$0.17\pm 0.05$ &\ldots &\ldots &\ldots &\ldots &\ldots &\ldots &$0.235\pm 0.008$ &\ldots &\ldots\\
$0.040$ &$0.26\pm 0.03$ &\ldots &\ldots &\ldots &\ldots &\ldots &\ldots &$0.252\pm 0.007$ &\ldots &\ldots\\
$0.050$ &$0.31\pm 0.03$ &\ldots &\ldots &\ldots &\ldots &\ldots &\ldots &$0.262\pm 0.005$ &\ldots &\ldots\\
$0.063$ &$0.31\pm 0.03$ &\ldots &$0.179\pm 0.005$ &$0.226\pm 0.006$ &$0.250\pm 0.006$ &\ldots &$0.34\pm 0.01$ &$0.258\pm 0.004$ &\ldots &\ldots\\
$0.079$ &$0.31\pm 0.02$ &\ldots &$0.192\pm 0.005$ &$0.268\pm 0.005$ &$0.267\pm 0.007$ &$0.201\pm 0.002$ &$0.369\pm 0.009$ &$0.254\pm 0.004$ &$0.128\pm 0.002$ &$0.248\pm0.026$ \\
$0.10$ &$0.314\pm 0.010$ &\ldots &$0.204\pm 0.004$ &$0.26\pm 0.03$ &$0.272\pm 0.006$ &$0.15\pm 0.02$ &$0.344\pm 0.009$ &$0.248\pm 0.003$ &$0.166\pm 0.002$ &$0.245\pm0.024$\\
$0.12$ &$0.314\pm 0.008$ &$0.188\pm 0.003$ &$0.197\pm 0.009$ &$0.31\pm 0.02$ &$0.261\pm 0.007$ &$0.18\pm 0.02$ &$0.321\pm 0.007$ &$0.249\pm 0.002$ &$0.166\pm 0.002$ &$0.243\pm0.021$\\
$0.14$ &$0.31\pm 0.02$ &$0.198\pm 0.003$ &$0.208\pm 0.006$ &$0.29\pm 0.02$ &$0.250\pm 0.006$ &$0.16\pm 0.02$ &$0.319\pm 0.007$ &$0.255\pm 0.002$ &$0.170\pm 0.009$ &$0.239\pm0.019$\\
$0.16$ &$0.28\pm 0.05$ &$0.202\pm 0.005$ &$0.200\pm 0.005$ &$0.27\pm 0.03$ &$0.250\pm 0.006$ &$0.18\pm 0.02$ &$0.317\pm 0.007$ &$0.259\pm 0.002$ &$0.18\pm 0.01$ &$0.237\pm0.017$\\
$0.19$ &$0.29\pm 0.04$ &$0.203\pm 0.004$ &$0.192\pm 0.006$ &$0.27\pm 0.02$ &$0.243\pm 0.008$ &$0.17\pm 0.02$ &$0.327\pm 0.006$ &$0.26\pm 0.01$ &$0.18\pm 0.01$ &$0.236\pm0.018$\\
$0.23$ &$0.26\pm 0.03$ &$0.210\pm 0.002$ &$0.186\pm 0.005$ &$0.24\pm 0.02$ &$0.245\pm 0.004$ &$0.18\pm 0.01$ &$0.338\pm 0.006$ &$0.258\pm 0.008$ &$0.154\pm 0.009$ &$0.230\pm0.018$\\
$0.27$ &$0.26\pm 0.03$ &$0.202\pm 0.004$ &$0.184\pm 0.005$ &$0.24\pm 0.02$ &$0.237\pm 0.004$ &$0.21\pm 0.01$ &$0.339\pm 0.004$ &$0.25\pm 0.01$ &$0.152\pm 0.008$ &$0.230\pm0.017$\\
$0.32$ &$0.25\pm 0.03$ &$0.20\pm 0.01$ &$0.170\pm 0.003$ &$0.24\pm 0.02$ &$0.225\pm 0.004$ &$0.21\pm 0.01$ &$0.28\pm 0.03$ &$0.248\pm 0.008$ &$0.149\pm 0.008$ &$0.219\pm0.014$\\
$0.37$ &$0.25\pm 0.03$ &$0.22\pm 0.02$ &$0.176\pm 0.003$ &$0.26\pm 0.02$ &$0.213\pm 0.004$ &$0.20\pm 0.01$ &$0.29\pm 0.03$ &$0.22\pm 0.01$ &$0.137\pm 0.009$ &$0.219\pm0.016$ \\
$0.44$ &$0.26\pm 0.03$ &$0.22\pm 0.02$ &$0.169\pm 0.003$ &$0.28\pm 0.02$ &$0.200\pm 0.003$ &$0.21\pm 0.01$ &$0.29\pm 0.03$ &$0.201\pm 0.009$ &$0.14\pm 0.01$ &$0.219\pm0.017$ \\
$0.52$ &$0.26\pm 0.02$ &$0.20\pm 0.02$ &$0.150\pm 0.004$ &$0.23\pm 0.03$ &$0.191\pm 0.003$ &$0.25\pm 0.02$ &$0.29\pm 0.03$ &$0.187\pm 0.008$ &$0.136\pm 0.007$ &$0.211\pm0.017$ \\
$0.61$ &$0.28\pm 0.02$ &$0.22\pm 0.01$ &$0.138\pm 0.004$ &$0.15\pm 0.04$ &$0.180\pm 0.006$ &$0.21\pm 0.02$ &$0.30\pm 0.03$ &$0.17\pm 0.01$ &$0.146\pm 0.009$ &$0.198\pm0.019$ \\
$0.72$ &$0.27\pm 0.02$ &$0.23\pm 0.02$ &\ldots &$0.082\pm 0.045$ &$0.154\pm 0.004$ &\ldots &$0.28\pm 0.03$ &$0.132\pm 0.008$ &$0.19\pm 0.01$ &$0.190\pm0.028$ \\
$0.85$ &$0.25\pm 0.02$ &$0.24\pm 0.03$ &\ldots &\ldots &\ldots &\ldots &$0.31\pm 0.03$ &$0.12\pm 0.01$ &\ldots &\ldots\\
$1.0$ &\ldots &$0.20\pm 0.03$ &\ldots &\ldots &\ldots &\ldots &$0.25\pm 0.04$ &\ldots &\ldots &\ldots\\
\enddata
\end{deluxetable}
\clearpage
\end{landscape}


\begin{thebibliography}{}
\bibitem[Allen et al.(2007)]{alle07} Allen, S.~W., Rapetti,
D.~A.,Schmidt, R.~W., Ebeling, H., Morris, R.~G., Fabian, A.~C. 2007,
\mnras, in press (arXiv.0706.0033)

\bibitem[B{\^i}rzan et al.(2004)]{2004ApJ...607..800B} B{\^i}rzan, L., 
Rafferty, D.~A., McNamara, B.~R., Wise, M.~W., \& Nulsen, P.~E.~J.\ 2004, 
\apj, 607, 800 

\bibitem[Binney \& Tremaine(1987)]{1987gady.book.....B} Binney, J., \& 
Tremaine, S.\ 1987, Princeton, NJ, Princeton University Press, 1987

\bibitem[Brighenti 
\& Mathews(1996)]{1996ApJ...470..747B} Brighenti, F., \& Mathews, W.~G.\ 1996, \apj, 470, 747 

\bibitem[Br{\"u}ggen et al.(2005)]{2005ApJ...628..153B} Br{\"u}ggen, M., 
Hoeft, M., \& Ruszkowski, M.\ 2005, \apj, 628, 153 

\bibitem[Brunetti \& Lazarian(2007)]{2007MNRAS.378..245B} Brunetti, G., \& 
Lazarian, A.\ 2007, \mnras, 378, 245 

\bibitem[Buote et al.(2007)]{2007ApJ...664..123B} Buote, D.~A., 
Gastaldello, F., Humphrey, P.~J., Zappacosta, L., Bullock, J.~S., 
Brighenti, F., \& Mathews, W.~G.\ 2007, \apj, 664, 123 

\bibitem[Buote et al.(2002)]{2002ApJ...577..183B} Buote, D.~A., Jeltema, 
T.~E., Canizares, C.~R., \& Garmire, G.~P.\ 2002, \apj, 577, 183 

\bibitem[Buote \& Canizares(1998)]{1998MNRAS.298..811B} Buote, D.~A., \& Canizares, C.~R.\ 1998, \mnras, 298, 811 

\bibitem[Buote \& Canizares(1996)]{1996ApJ...457..565B} Buote, D.~A., \& Canizares, C.~R.\ 1996, \apj, 457, 565 

\bibitem[Buote \& Canizares(1994)]{1994ApJ...427...86B} Buote, D.~A., \& Canizares, C.~R.\ 1994, \apj, 427, 86 

\bibitem[Buote \& Tsai(1996)]{1996ApJ...458...27B} Buote, D.~A., \& Tsai, 
J.~C.\ 1996, \apj, 458, 27 

\bibitem[Buote \& Tsai(1995)]{1995ApJ...452..522B} Buote, D.~A., \& Tsai, 
J.~C.\ 1995, \apj, 452, 522 

\bibitem[Burns et al.(2007)]{2007arXiv0708.1954B} Burns, J.~O., Hallman, 
E.~J., Gantner, B., Motl, P.~M., \& Norman, M.~L.\ 2007, ArXiv e-prints, 
708, arXiv:0708.1954 

\bibitem[Carter \& Metcalfe(1980)]{1980MNRAS.191..325C} Carter, D., \& 
Metcalfe, N.\ 1980, \mnras, 191, 325 

\bibitem[Chepurnov \& Lazarian(2006)]{2006astro.ph.11463C} Chepurnov, A., 
\& Lazarian, A.\ 2006, ArXiv Astrophysics e-prints, arXiv:astro-ph/0611463 

\bibitem[David et al.(2001)]{2001ApJ...557..546D} David, L.~P., Nulsen, 
P.~E.~J., McNamara, B.~R., Forman, W., Jones, C., Ponman, T., Robertson, 
B., \& Wise, M.\ 2001, \apj, 557, 546 

\bibitem[Dolag et al.(2005)]{2005MNRAS.364..753D} Dolag, K., Vazza, F., 
Brunetti, G., \& Tormen, G.\ 2005, \mnras, 364, 753 

\bibitem[Dupke \& Bregman(2006)]{2006ApJ...639..781D} Dupke, R.~A., \& 
Bregman, J.~N.\ 2006, \apj, 639, 781 

\bibitem[Evrard et al.(1996)]{1996ApJ...469..494E} Evrard, A.~E., Metzler, 
C.~A., \& Navarro, J.~F.\ 1996, \apj, 469, 494 

\bibitem[Fabian et al.(2006)]{2006MNRAS.366..417F} Fabian, A.~C., Sanders, 
J.~S., Taylor, G.~B., Allen, S.~W., Crawford, C.~S., Johnstone, R.~M., \& 
Iwasawa, K.\ 2006, \mnras, 366, 417 

\bibitem[Faltenbacher et al.(2005)]{2005MNRAS.358..139F} Faltenbacher, A., 
Kravtsov, A.~V., Nagai, D., \& Gottl{\"o}ber, S.\ 2005, \mnras, 358, 139 

\bibitem[Finoguenov et al.(2001)]{2001ApJ...555..191F} Finoguenov, A., 
Arnaud, M., \& David, L.~P.\ 2001, \apj, 555, 191 

\bibitem[Forman et al.(2005)]{2005ApJ...635..894F} Forman, W., et al.\ 
2005, \apj, 635, 894 

\bibitem[Gastaldello et al.(2007)]{2007ApJ...669..158G} Gastaldello, F., 
Buote, D.~A., Humphrey, P.~J., Zappacosta, L., Bullock, J.~S., Brighenti, 
F., \& Mathews, W.~G.\ 2007, \apj, 669, 158 

\bibitem[Hallman et al.(2006)]{2006ApJ...648..852H} Hallman, E.~J., Motl, 
P.~M., Burns, J.~O., \& Norman, M.~L.\ 2006, \apj, 648, 852 


\bibitem[Inogamov \& Sunyaev(2003)]{2003AstL...29..791I} Inogamov, N.~A., 
\& Sunyaev, R.~A.\ 2003, Astronomy Letters, 29, 791 

\bibitem[Jeltema et al.(2008)]{2008ApJ...681..167J} Jeltema, T.~E., 
Hallman, E.~J., Burns, J.~O., \& Motl, P.~M.\ 2008, \apj, 681, 167 

\bibitem[Kay et al.(2004)]{2004MNRAS.355.1091K} Kay, S.~T., Thomas, P.~A., 
Jenkins, A., \& Pearce, F.~R.\ 2004, \mnras, 355, 1091 

\bibitem[Kley 
\& Mathews(1995)]{1995ApJ...438..100K} Kley, W., \& Mathews, W.~G.\ 1995, \apj, 438, 100 

\bibitem[Klypin et al.(2001)]{2001ApJ...554..903K} Klypin, A., Kravtsov, 
A.~V., Bullock, J.~S., \& Primack, J.~R.\ 2001, \apj, 554, 903 

\bibitem[Kravtsov et al.(2006)]{2006ApJ...650..128K} Kravtsov, A.~V., 
Vikhlinin, A., \& Nagai, D.\ 2006, \apj, 650, 128 

\bibitem[Kravtsov et al.(2002)]{2002ApJ...571..563K} Kravtsov, A.~V., 
Klypin, A., \& Hoffman, Y.\ 2002, \apj, 571, 563 

\bibitem[Lau et al.(2008)]{} Lau, E.\ 2008, in preparation

\bibitem[Lewis et al.(2003)]{2003ApJ...586..135L} Lewis, A.~D., Buote, 
D.~A., \& Stocke, J.~T.\ 2003, \apj, 586, 135 

\bibitem[Mathews 
\& Brighenti(2003)]{2003ARA&A..41..191M} Mathews, W.~G., \& Brighenti, F.\ 2003, \araa, 41, 191 

\bibitem[Nagai et al.(2007)]{2007ApJ...655...98N} Nagai, D., Vikhlinin, A., 
\& Kravtsov, A.~V.\ 2007, \apj, 655, 98 

\bibitem[Nulsen et al.(1984)]{1984MNRAS.208..185N} Nulsen, P.~E.~J., 
Stewart, G.~C., \& Fabian, A.~C.\ 1984, \mnras, 208, 185 

\bibitem[Pawl et al.(2005)]{2005ApJ...631..773P} Pawl, A., Evrard, A.~E., 
\& Dupke, R.~A.\ 2005, \apj, 631, 773 

\bibitem[Pointecouteau et al.(2005)]{2005A&A...435....1P} Pointecouteau, E., Arnaud, M., \& Pratt, G.~W.\ 2005, \aap, 435, 1 

\bibitem[Pratt \& Arnaud(2003)]{2003A&A...408....1P} Pratt, G.~W., \& Arnaud, M.\ 2003, \aap, 408, 1 

\bibitem[Rasia et al.(2006)]{2006MNRAS.369.2013R} Rasia, E., et al.\ 2006, 
\mnras, 369, 2013 


\bibitem[Rasia et al.(2004)]{2004MNRAS.351..237R} Rasia, E., Tormen, G., \& 
Moscardini, L.\ 2004, \mnras, 351, 237 

\bibitem[Sanderson et al.(2006)]{2006MNRAS.372.1496S} Sanderson, A.~J.~R., 
Ponman, T.~J., \& O'Sullivan, E.\ 2006, \mnras, 372, 1496 

\bibitem[Schuecker et al.(2004)]{2004A&A...426..387S} Schuecker, P., 
Finoguenov, A., Miniati, F., B{\"o}hringer, H., \& Briel, U.~G.\ 2004, 
\aap, 426, 387 

\bibitem[Stanek et al.(2004)]{stan06} Stanek, R., Evrard, A.~E.,
B{\"o}hringer, H., Schuecker, P. \& Nord, B.\ 2006, \apj, 648, 956

\bibitem[Sunyaev et al.(2003)]{2003AstL...29..783S} Sunyaev, R.~A., Norman, 
M.~L., \& Bryan, G.~L.\ 2003, Astronomy Letters, 29, 783 


\bibitem[Thomas et al.(1998)]{1998MNRAS.296.1061T} Thomas, P.~A., et al.\ 
1998, \mnras, 296, 1061 

\bibitem[Tsai et al.(1994)]{1994ApJ...423..553T} Tsai, J.~C., Katz, N., \& 
Bertschinger, E.\ 1994, \apj, 423, 553 


\bibitem[Vazza et al.(2006)]{2006MNRAS.369L..14V} Vazza, F., Tormen, G., 
Cassano, R., Brunetti, G., \& Dolag, K.\ 2006, \mnras, 369, L14 

\bibitem[Vikhlinin et al.(2006)]{2006ApJ...640..691V} Vikhlinin, A., 
Kravtsov, A., Forman, W., Jones, C., Markevitch, M., Murray, S.~S., 
\& Van Speybroeck, L.\ 2006, \apj, 640, 691 

\bibitem[Vikhlinin et al.(2005)]{2005ApJ...628..655V} Vikhlinin, A., 
Markevitch, M., Murray, S.~S., Jones, C., Forman, W., \& Van Speybroeck, 
L.\ 2005, \apj, 628, 655 


\bibitem[Zappacosta et al.(2006)]{2006ApJ...650..777Z} Zappacosta, L., 
Buote, D.~A., Gastaldello, F., Humphrey, P.~J., Bullock, J., Brighenti, F., 
\& Mathews, W.\ 2006, \apj, 650, 777 

\end{thebibliography}
\end{document}